\documentclass[12pt]{article}
\pdfoutput=1
\usepackage{amssymb}
\usepackage{amsmath}
\usepackage{latexsym}
\usepackage[usenames]{color}
\usepackage{cite}
\definecolor{darkblue}{cmyk}{0.9,0.9,0,0}
\definecolor{darkred}{rgb}{0.6,0,0.3}
\usepackage{graphicx} 
\usepackage[setpagesize=false,pagebackref=false, linktocpage, bookmarksopen=true, colorlinks=true, linkcolor=darkblue,citecolor=darkblue,urlcolor=darkblue]{hyperref}
\usepackage{hyperref}
\newcommand{\arXiv}[2]{\href{http://arxiv.org/abs/#1}{{\tt arXiv:#2}}}
\newcommand{\hep}[2]{\href{http://arxiv.org/abs/#1}{{\tt #2}}}

\renewcommand{\thefootnote}{\arabic{footnote}}
\allowdisplaybreaks

\def\fn#1{\footnote{#1}}

\def\eqref#1{(\ref{#1})}
\def\comma{\,,}
\def\period{\,.}

\newcommand{\beq}{\begin{equation}}
\newcommand{\eeq}{\end{equation}}

\begin{document}
\thispagestyle{empty}

\renewcommand{\thefootnote}{\fnsymbol{footnote}}
\setcounter{page}{1}
\setcounter{footnote}{0}
\setcounter{figure}{0}
\begin{center}
$$$$
{\Large\textbf{\mathversion{bold}
Hexagonalization of Correlation Functions II : \\Two-Particle
Contributions
}\par}

\vspace{1.3cm}

\textrm{Thiago Fleury$^{\textcolor[rgb]{0,0.6,0}{\blacktriangledown}}$, Shota Komatsu$^{\textcolor[rgb]{1,0.8,0}{\blacktriangleright}}$}
\\ \vspace{1.2cm}
\footnotesize{\textit{
$^{\textcolor[rgb]{0,0.6,0}{\blacktriangledown}}$ 
Instituto de F\'isica Te\'orica, UNESP - Univ. Estadual Paulista, 
ICTP South American Institute for Fundamental Research,
Rua Dr. Bento Teobaldo Ferraz 271, 01140-070, S\~ao Paulo, SP, Brasil
\vspace{1mm} \\
$^{\textcolor[rgb]{0,0.6,0}{\blacktriangledown}}$ 
Laboratoire de Physique Th\'eorique
de l'Ecole Normale Sup\'erieure et l'Universit\'e Paris-VI, 24 rue Lhomond, Paris CEDEX 75231, France \\
$^{\textcolor[rgb]{0,0.6,0}{\blacktriangledown}}$ 
International Institute of Physics, Federal University of Rio Grande do Norte, \\ Campus Universit\'ario,  Lagoa Nova, Natal, RN 59078-970, Brazil
 \\  
\vspace{0.3mm}
$^{\textcolor[rgb]{1,0.8,0}{\blacktriangleright}}$  School of Natural Sciences, Institute for
Advanced Study, Princeton, NJ 08540, USA\\
}  
\vspace{4mm}
}

\par\vspace{1.5cm}

\textbf{Abstract}\vspace{2mm}
\end{center}

In this work, we compute one-loop planar 
five-point functions in $\mathcal{N}$=4 super-Yang-Mills
using integrability. As in the previous work, we 
decompose
the correlation functions  into
hexagon form factors and glue them using the 
weight factors which depend on the cross-ratios. The 
main new ingredient in the computation, as compared 
to the four-point functions studied in the previous paper, is the two-particle mirror contribution. We develop techniques to evaluate it and find agreement with the perturbative results in all the cases we analyzed.  
In addition, we consider next-to-extremal four-point 
functions, which are known to be protected, and 
show that the sum of one-particle and two-particle 
contributions 
at one loop adds up to zero as expected.
The tools developed in this work would be useful for 
computing higher-particle contributions which would 
be relevant for more complicated quantities such 
as higher-loop corrections and non-planar correlators.

\noindent

\setcounter{page}{1}
\renewcommand{\thefootnote}{\arabic{footnote}}
\setcounter{footnote}{0}
\setcounter{tocdepth}{2}
\newpage
\tableofcontents

\parskip 5pt plus 1pt   \jot = 1.5ex

\section{Introduction\label{sec:intro}}

The planar $\mathcal{N}$=4 super-Yang-Mills is believed 
to be an integrable theory \cite{review} to all orders in perturbation
theory\fn{For more recent progress, in particular about the quantum spectral curve method, see the original article \cite{QuantumSpectralCurve}  and also the lecture notes \cite{GromovIntroduction}.}. Recently, an integrability-based method for computing the 
four- \cite{EdenSfrondrini, FleuryKomatsu} and higher-point \cite{FleuryKomatsu} correlation functions, named hexagonalization, was proposed. 
The method consists in cutting the 
correlation functions into 
smaller building blocks called the hexagon form factors which were first introduced in \cite{BKV} in the context of the three-point
functions. To be more concrete, take for example a planar $n$-point function. Pictorially, it
can be represented as a sphere with $n$ holes and 
one cuts this surface into $2 (n-2)$ hexagonal patches.
The contribution from each patch is given by the hexagon form factor, and by gluing these hexagons with appropriate weight factor, one can compute the original correlation function\fn{Alternatively, one could perform the OPE and then compute the structure constants using the integrability \cite{AsymptoticFour}. However
this OPE method involves the mixing with multi-trace operators
and it is not clear how to study this mixing systematically.
An advantage of using the hexagonalization is that it avoids the mixing problem.
It would be very interesting to understand the
relation between the two approaches. 
}.

In \cite{FleuryKomatsu}, we applied this method to compute four-point functions at one loop. In this work, we generalize this analysis to some of the one-loop five-point functions. For the sake of simplicity, we consider a restricted kinematics in which all five operators live on a  two-dimensional plane\fn{We also impose analogous constraints on the $R$-charge polarizations.}. From the computational point of view, the main difference from the previous work is that we now need the two-particle mirror contributions while, in all the calculations done in \cite{FleuryKomatsu}, only the one-particle contribution was needed. We compare our results with the perturbative data \cite{DrukkerPlefka} and show that they agree. 
We also show that the integrability result is independent of
the way of cutting the worldsheet into hexagons. This property was named the flip invariance in \cite{FleuryKomatsu}. The way the flip invariance is realized in the five-point functions is much more nontrivial as compared to that for the four-point functions, and it serves as a stringent consistency check of our computation. 

The tools developed in this paper should also be useful for computing higher-particles contributions, which would be relevant for higher-point functions and nonplanar correction \cite{nonplanarpaper,nonplanarpaper2}\fn{It is also interesting to study the non-planar correction to the non-BPS two-point functions. The initial attempt in this direction was made in \cite{EdenNew}. The computation is however generally more involved and our tools do not immediately apply to such cases since one also needs to include physical magnons.}.
As another application of our result, we provide supporting evidence for the prescription conjectured in \cite{FleuryKomatsu}, which states that one only needs to consider the mirror corrections coming from {\it{1-edge irreducible}} (1EI) graphs, i.e. graphs that are still connected when any bridge with non-zero length is cut. Specifically,
we consider one-loop next-to-extremal four-point functions, which consist only of non-1EI graphs, and show that the mirror particle corrections for such correlators cancel among themselves. This is in accordance with the non-renormalization property of the next-to-extremal correlators discussed in \cite{Extremal2,Extremal3,Extremal4}.

The paper is organized as follows. 
In section \ref{sec:5pt}, we present the prediction from integrability and compare it with the one-loop perturbative data for five-point functions. In all the examples that we studied, we find a perfect match. We then sketch the strategy of the computation of two-particle contributions in section \ref{sec:2particle} relegating more technical details to the appendices. 
The flip invariance is also discussed in great detail in section 
\ref{sec:2particle}. 
Using the two-particle contribution, we furthermore show the cancellation of the mirror-particle 
corrections coming from non-1EI graphs for a near
extremal four-point function in section \ref{TheFate1EI}.
Section \ref{sec:conclusion} has our conclusions. Appendices are basically for explaining the technical details of the computation: Appendix \ref{ZmarkersSection} has the $Z$-marker prescription which we used to dress
the mirror bound state basis. The mirror bound state $S$-matrix 
is computed in appendix \ref{ap:S-matrix}. The necessary weak coupling
expansions are shown in appendix \ref{WeakCouplingExpansions}. The complete 
two-particle integrand is given in appendix \ref{TheIntegrand}. There, we also comment on how to evaluate the integral and the prescription to avoid singularities.

\section{Five-Point Functions: Perturbation and Integrability}\label{sec:5pt}

In this section, we compare the integrability result and the perturbative data for five-point functions at one loop.
Unlike the four-point functions where a simple closed-form expression
is known for BPS operators of any length, such an expression is not known for the five-point functions\fn{In principle, one can use the method of \cite{DrukkerPlefka} to compute any desired five-point functions. However, a closed-form expression similar to the ones given in \cite{AllThreeLoop} is not known.}. 
We are thus going to focus on two examples,
the case of five length-two operators and the case of three length-two 
and two length-three operators, which were computed and written down explicitly in \cite{DrukkerPlefka}. Another reason why we focus on these two examples is because they receive corrections only from the one- and the two-particle contributions. On the other hand, many other examples such as the correlator of four length-two and one length-four operators needs higher-particle contributions. Although we do not perform the computation of such contributions in this paper, in principle they can be 
calculated using the tools developed in this paper. 
\subsection{Set-up}
We consider BPS operators and we denote them as 
\begin{equation}
\mathcal{O}_{L_i} (x_i, Y_i) = {\rm{Tr}} \, ( (Y_i \cdot \Phi)^{L_i} (x_i)) \, , 
\end{equation} 
where $Y_i \cdot \Phi =\sum_{I=1}^6 Y_i^I \Phi^I$. $Y_i$'s are null polarization vectors
$Y_i \cdot Y_i =0 $, $\Phi^I$'s are the six scalars
and
$L_i$ is the length of the operator.
We normalize the two-point functions of these operators as      
\begin{equation}
\langle \mathcal{O}_{L}(x_1,Y_1) \mathcal{O}_{L}(x_2,Y_2) \rangle = 
( d_{12} )^{L} \, ,   
\end{equation}
with
\begin{equation}
\label{eq:definitionofd}
d_{ij} = \frac{y^2_{ij}}{x_{ij}^2} \, , \quad {\rm{and}} \quad 
y_{ij}^2 = Y_i \cdot Y_j \, .  
\end{equation}

In our normalization, the planar five-point functions of such BPS operators are of the form,
\begin{equation}\label{connecteddef}
\begin{aligned}
\langle \mathcal{O}_{L_1}(x_1,Y_1) 
\mathcal{O}_{L_2}(x_2,Y_2)
\mathcal{O}_{L_3}(x_3,Y_3)
\mathcal{O}_{L_4}(x_4,Y_4)
\mathcal{O}_{L_5}(x_5,Y_5) \rangle=  \\
\; ( {\rm{disconnected}} ) 
+ \frac{\prod_{i=1}^{5} \sqrt{L_i}}{N^{3}} \, G_{\{L_i \}} \, ,\nonumber
\hspace{25mm} 
\end{aligned}
\end{equation}
where the first term on the right hand side denotes the disconnected part of the correlator which is given by a product of lower-point functions.
In this work, we are only interested in the connected correlator 
$G_{\{L_i \}}$ up to one-loop order. 

For the sake of simplicity, we are going to work with a restricted kinematics, i.e. we
consider the configurations in which the five operators live in a plane 
both in spacetime and in $R$-charge. This reduces the number of cross ratios from ten (five for spacetime and five for $R$-charge) to eight (four for spacetime and four for $R$-charge). 
They are defined as   
\beq
\begin{aligned}
\label{eq:TheCrossRatios}
&z \bar{z}=\frac{x_{12}^2x_{34}^2}{x_{13}^2x_{24}^2}\comma 
\quad (1-z)(1-\bar{z})=\frac{x_{14}^2x_{23}^2}{x_{13}^2x_{24}^2}\comma 
\quad \alpha\bar{\alpha}=\frac{y_{12}^2y_{34}^2}{y_{13}^2y_{24}^2}\comma
\quad (1-\alpha)(1-\bar{\alpha})=\frac{y_{14}^2y_{23}^2}{y_{13}^2y_{24}^2}
\comma\\
&w \bar{w}=\frac{x_{15}^2x_{34}^2}{x_{13}^2x_{54}^2}\comma \quad 
(1-w)(1-\bar{w})=\frac{x_{14}^2x_{35}^2}{x_{13}^2x_{45}^2}\comma 
\quad \beta\bar{\beta}=\frac{y_{15}^2y_{34}^2}{y_{13}^2y_{54}^2}\comma
\quad (1-\beta)(1-\bar{\beta})=\frac{y_{14}^2y_{35}^2}{y_{13}^2y_{45}^2} \, . 
\end{aligned}
\eeq
Because of the restricted kinematics, other cross ratios can be written in terms of the aforementioned ones as 
\begin{equation}
\frac{x_{13}^2x_{52}^2}{x_{15}^2x_{32}^2} = 
\frac{(z - w)(\bar{z}- \bar{w})}{ w \bar{w} \, (z- 1)(\bar{z}-1)} \, ,  
\quad {\rm{and}} \quad \frac{y_{13}^2 y_{52}^2}{y_{15}^2 y_{32}^2} = 
\frac{(\alpha - \beta)(\bar{\alpha}- \bar{\beta})}
{ \beta \bar{\beta} \, (\alpha-1)(\bar{\alpha}-1)} \, . 
\end{equation}
As will be explained more in detail in section \ref{sec:2particle}, this restriction simplifies the computation of the two-particle contribution since we only need diagonal components of the matrix part of the hexagon form factor. On the other hand, general kinematics necessitates non-diagonal components since the generators that take the operators away from the plane have to be included in the weight factor. We wish to emphasize, however, that this is just a technical problem rather than a conceptual problem, and could be overcome by generalizing the analysis in this paper. We leave it for future investigations.

\subsection{Results from Integrability}
We now summarize the basic building blocks for the integrability results. The details of the computation will be explained in section \ref{sec:2particle}.

\begin{figure}[t]
\centering
\begin{minipage}{0.3\hsize}
\centering
\includegraphics[clip,height=4cm]{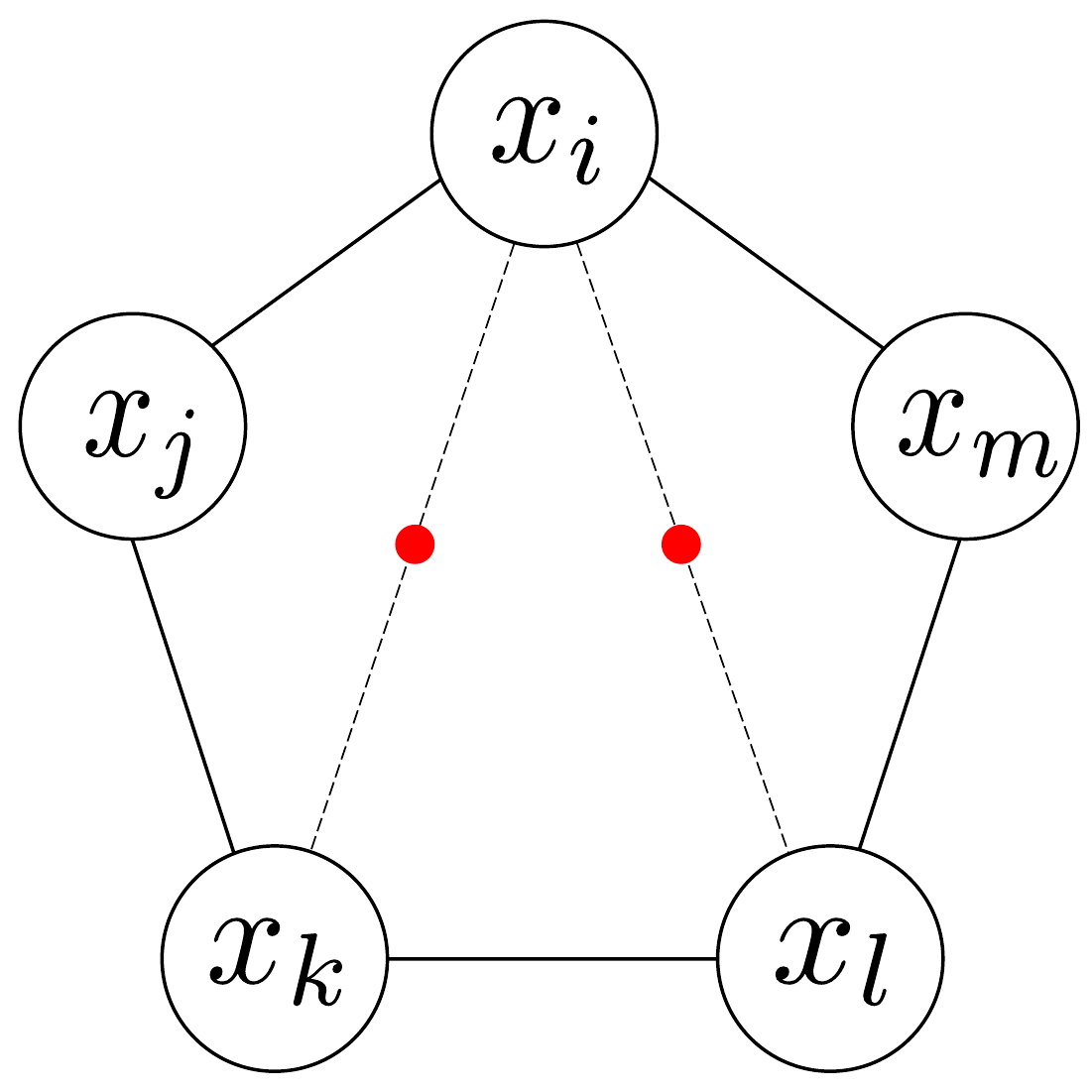} 
\end{minipage}
\begin{minipage}{0.3\hsize}
\centering
\includegraphics[clip,height=4cm]{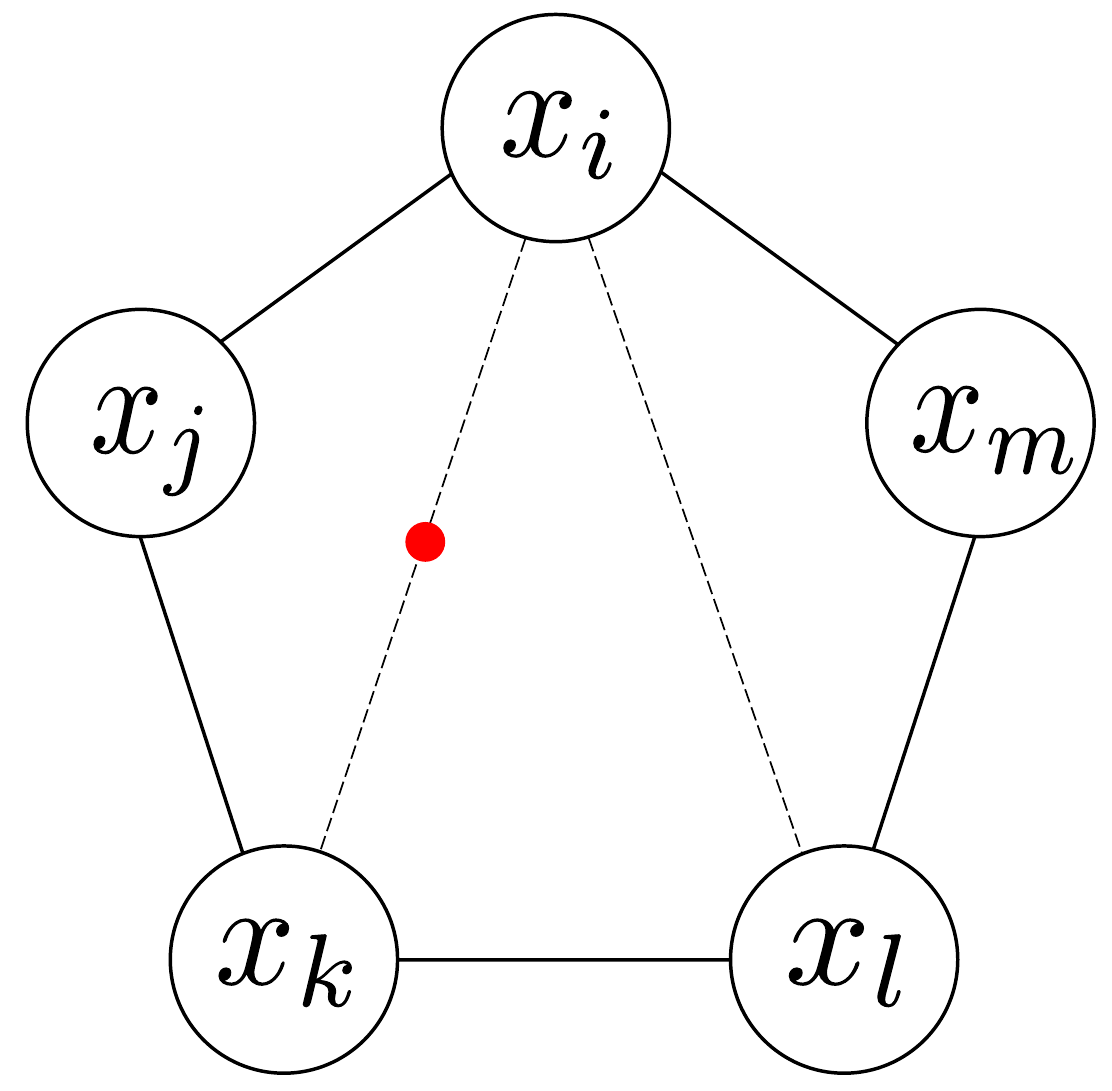} 
\end{minipage}
\begin{minipage}{0.3\hsize}
\centering
\includegraphics[clip,height=4cm]{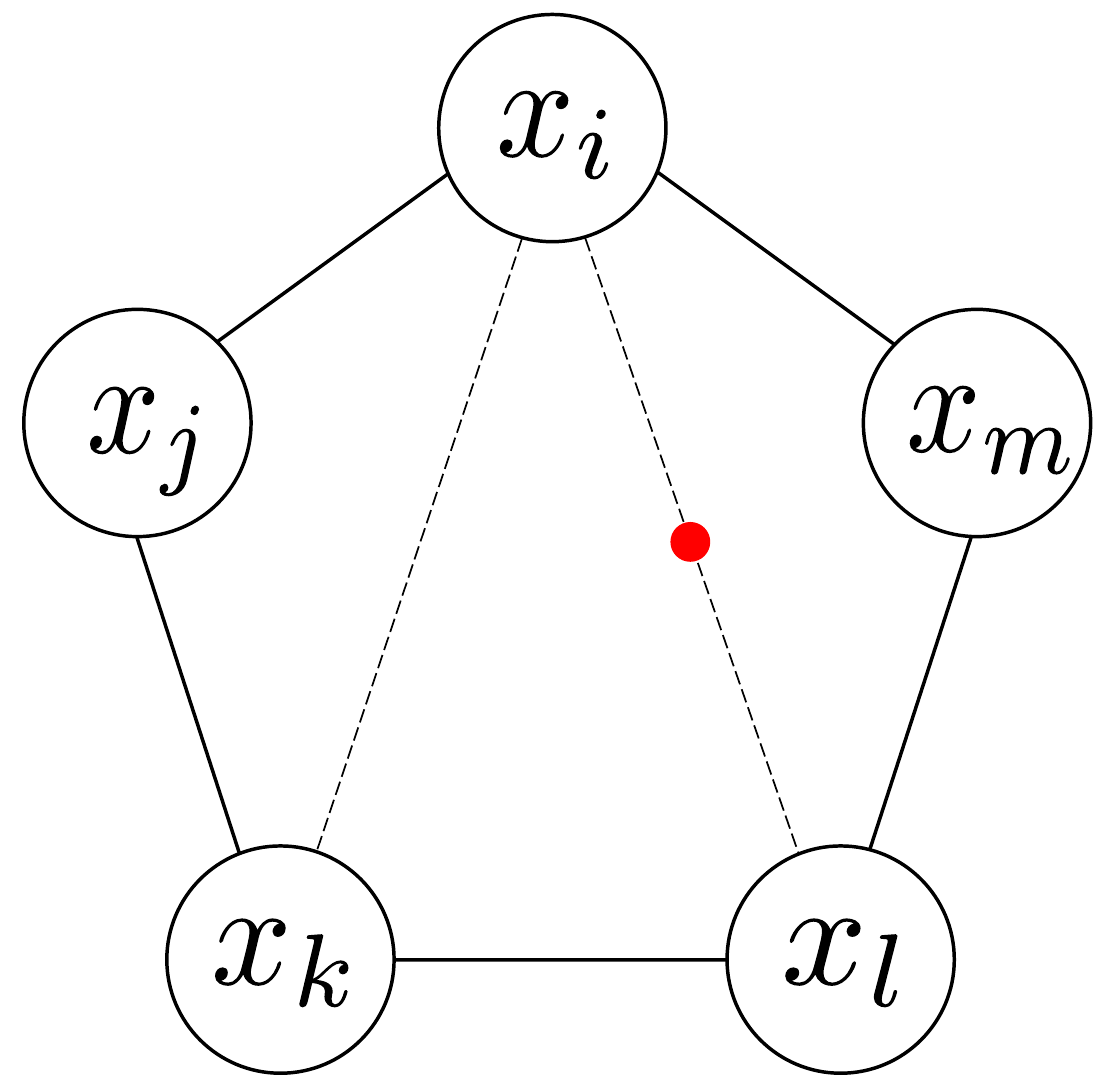} 
\end{minipage}\\
\begin{minipage}{0.3\hsize}
\centering
$(a)$
\end{minipage}
\begin{minipage}{0.3\hsize}
\centering
$(b)$
\end{minipage}
\begin{minipage}{0.3\hsize}
\centering
$(c)$
\end{minipage}
\caption{The two-particle mirror correction, $(a)$, and the related one-particle corrections, $(b)$ and $(c)$. The red dots are the mirror particles and the dashed lines are zero length
bridges. The sum of three contributions gives a one-loop correction to the decagon and is given by a simple combination of the one-loop conformal integrals as shown in \eqref{decagonsum}.} 
\label{fig:TheTwoMirrorParticlesFig}
\end{figure}

The integrability computation for the five-point 
functions considered in this work 
involves the one-particle and the two-particle
mirror contributions. More precisely, what we need are the contributions from an one-particle state living on a length-zero bridge and the contributions from two one-particle states living on neighbouring length-zero bridges. See figure \ref{fig:TheTwoMirrorParticlesFig}-$(a)$ for a pictorial explanation.

\begin{figure}[t]
\centering
\includegraphics[clip,height=4cm]{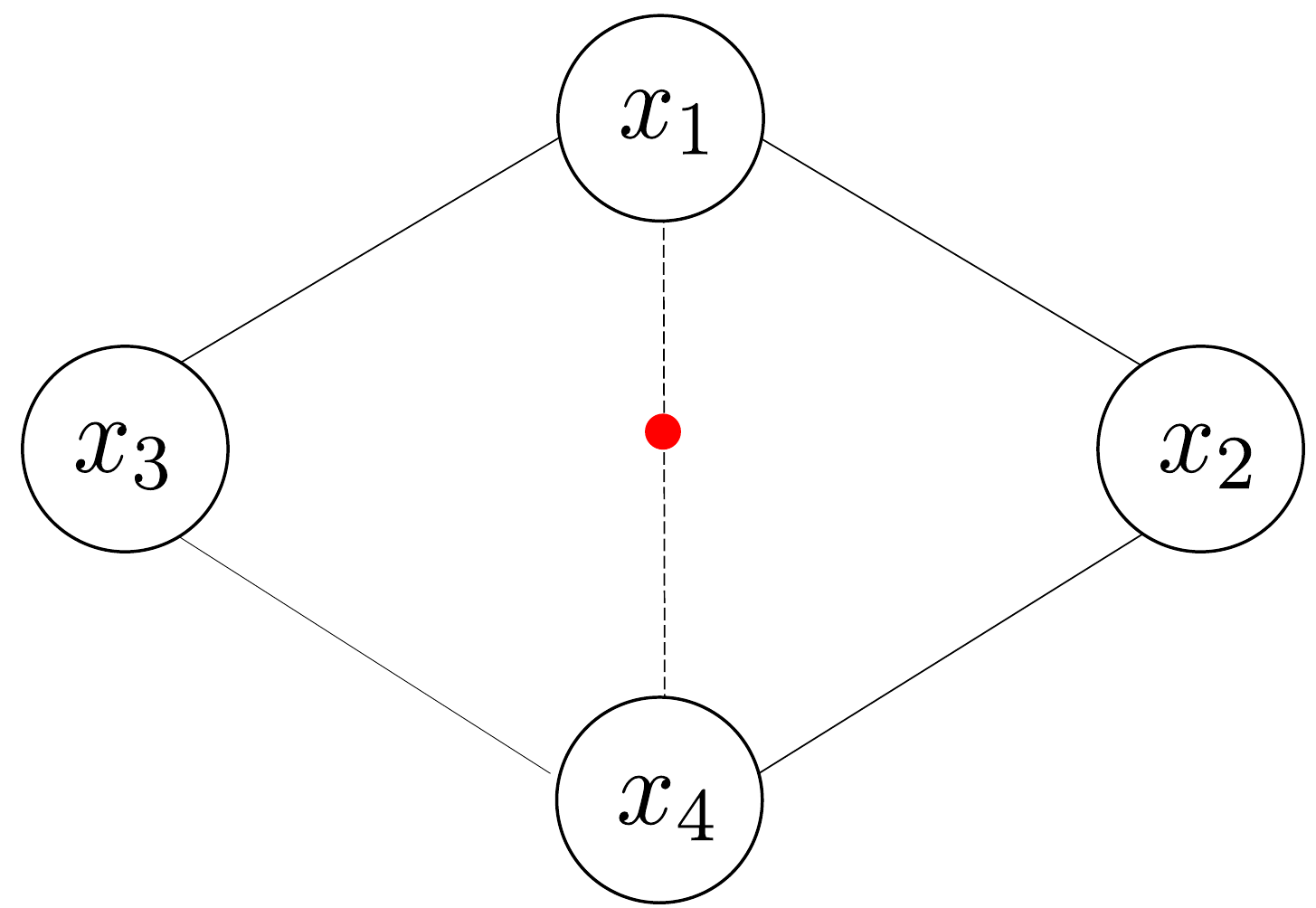}
\caption{The configuration for the one-particle mirror correction. The dashed line denotes the length-zero bridge while the red dot denotes the mirror particle. The result is given in \eqref{eq:oneparticle}.}
\label{fig:oneparticleconfig}
\end{figure}

The one-particle contribution was computed
in \cite{FleuryKomatsu}. For the configuration depicted in figure \ref{fig:oneparticleconfig}, the result can be written as\fn{This expression makes manifest the invariance under the flip transformation, $z\to z^{-1}$ and $\alpha \to \alpha^{-1}$.}     
\beq
\label{eq:oneparticle}
\mathcal{M}^{(1)}(z, \alpha) = m(z)+m(z^{-1}) \, , 
\eeq
where $m(z)$ is given by
\beq\label{mfunctiondef}
m(z)\equiv g^2\frac{(z+\bar{z})-(\alpha+\bar{\alpha})}{2}F^{(1)}(z,\bar{z})\comma
\eeq
and the cross ratios are defined in \eqref{eq:TheCrossRatios}. As is clear from the definition above, $m(z)$ is actually a function of four cross ratios $z$, $\bar{z}$, $\alpha$ and $\bar{\alpha}$. We however only write the dependence on the first argument $z$ since it is easy to figure out the dependence on the other arguments. $F^{(1)}$ and $g$ are defined by
\begin{equation}
\label{eq:ConformalIntegral}
F^{(1)}(z, \bar{z}) \equiv
\frac{\Phi (z,\bar{z})}{z-\bar{z}}  \quad
\left(= \frac{x_{13}^2 x_{24}^2}{\pi^2} \int \frac{d^4 x_5}
{x_{15}^2 x^2_{25} x^2_{35} x^2_{45}} \right) \, \quad {\rm{and}}
\quad  g^ 2 = \frac{\lambda}{ 16 \pi^ 2}  \, , 
\end{equation}
with $\lambda$ being the `t Hooft coupling.  The function $\Phi (z,\bar{z})$ is given by
\beq
\Phi (z,\bar{z})=2 {\rm Li}_{2} (z)-2 {\rm Li}_2 (\bar{z})+\log ( z \bar{z}) \log \left( \frac{1-z}{1-\bar{z}}\right)\period
\eeq
The function $m(z)$ has several important properties,
\beq\label{propertiesofmz}
\begin{aligned}
&m(0)=m(1)=m(\infty)=0\comma\\
&m(z)+m(1-z)=0\comma
\end{aligned}
\eeq
which we use often in this paper.

The computation of the two-particle contribution is the main outcome of this work  and will be explained in section \ref{sec:2particle}.
The result turns out to be given by a linear combination of one-loop conformal integrals. For the configuration depicted in figure \ref{fig:TheTwoMirrorParticlesFig}-$(a)$, the result reads
\begin{equation}
\label{eq:twoparticle}
\begin{aligned}
\mathcal{M}^{(2)}(z_1, z_2, \alpha_1, \alpha_2) = &-m(z_1)-m(z_2^{-1})\\
&+m\left(\frac{z_1-1}{z_1z_2}\right)+m\left(\frac{1-z_1+z_1z_2}{z_2}\right)+m\left(z_1(1-z_2)\right)\comma
\end{aligned}
\end{equation}
where 
the cross ratios in the formula are given by
\beq\label{crossratiorule}
z_1 \bar{z}_1 = 
\frac{x_{i m}^2 x_{k l}^2}{x_{i k}^2 x_{m l}^2} 
\, , \,\, (1-z_1)(1-\bar{z}_1)= \frac{x_{i l}^2 
x_{k m}^2}{
x_{i k }^2 x_{l m}^2} \, , \,\,  
z_2 \bar{z}_2 = 
\frac{x^2_{i l} x^2_{j k}}{x^2_{ij} x^2_{lk}} 
\, , \,\, (1-z_2)(1-\bar{z}_2)= 
\frac{x_{i k}^2 x_{j l}^2}{x^2_{i j}
x^2_{k l}} \, , 
\eeq
and similarly for the $R$-charge cross-ratios $\alpha_1$ and $\alpha_2$.   

One can then compute the sum of the one-particle and the two-particle corrections that contribute to the decagon depicted in figure \ref{fig:TheTwoMirrorParticlesFig} as follows:
\beq\label{decagonsum}
\begin{aligned}
&\mathcal{M}^{(1)}(z_1,\alpha_1)+\mathcal{M}^{(1)}(z_2,\alpha_2)+\mathcal{M}^{(2)}(z_1,z_2,\alpha_1,\alpha_2)=\\
&m(z_1^{-1})+m(z_2)+m\left(\frac{z_1-1}{z_1z_2}\right)+m\left(\frac{1-z_1+z_1z_2}{z_2}\right)+m\left(z_1(1-z_2)\right)\period
\end{aligned}
\eeq
An important property of this expression is that the five terms in the formula correspond to all possible cross ratios that can be formed inside the five-point graph depicted in figure \ref{fig:TheTwoMirrorParticlesFig}. This makes it clear that the result is independent of the way we cut the graph into hexagons. This property, called the flip invariance, will be discussed more in detail in section \ref{subsec:flipinvariance}.

In the following two subsections, we show that one can reproduce the perturbative data for five-point functions from these expressions.

\subsection{Comparison I: Five 20{$^{\prime}$}} 
Let us first study the simplest five-point functions, which is the correlator of five length-two operators (also known as $20^{\prime}$ operators).
 
To express the perturbative result, we introduce the definitions 
\begin{equation}
s_{ijkl} = \frac{x^2_{ij} x^2_{kl}}{x^2_{ik}x^2_{jl}}  \equiv
z_{i j k l} \, \bar{z}_{i j k l} \, , \quad \quad t_{i j k l} = \frac{x^2_{i l} x^2_{j k}}{x^2_{ik} 
x^2_{j l}} \equiv (1- z_{i j k l})(1- \bar{z}_{i j k l}) \, .  
\end{equation}
and 
\beq
\label{eq:TheDperturbative}
D_{ijkl} = g^2 F^{(1)}(z_{ijkl}, \bar{z}_{ijkl})
(2 d_{ik} d_{jl} + (s_{ijkl}-1-t_{ijkl})d_{il}d_{jk}
+(t_{ijkl}-1-s_{ijkl})d_{ij}d_{kl}) \, , 
\eeq
which can also be written in terms of $m(z)$ as
\beq
D_{ijkl}=2 m(z_{ijlk})d_{il}d_{jk}+2m(z_{iljk})d_{ij}d_{lk}\period
\eeq
Then the one-loop result in \cite{DrukkerPlefka} can be expressed as 
\begin{equation}
\label{eq:five20primedata}
\begin{aligned}
\left.G^{1}_{\{2,2,2,2,2\}}\right|_{\rm perturbation}= 
-&\left(D_{1234} [ 1 \, 3, 2 \, 4 | 5] + D_{1 3 2 4} [ 1 \, 2, 3 \, 4 | 5]
+ D_{1243} [ 1 \, 4, 2 \, 3 | 5] \right. \\
&+D_{1235} [ 1 \, 3, 2 \, 5 | 4] + D_{1325} [ 1 \, 2, 5 \, 3 | 4]
+ D_{1253} [ 1 \, 5, 2 \, 3 | 4] \\
&+D_{1254} [ 1 \, 5, 2 \, 4 | 3] + D_{1524}
[ 1 \, 2, 4 \, 5 | 3] + D_{1245} [ 1 \, 4, 2 \, 5 | 3] \\
&+D_{1534} [ 1 \, 3, 5 \, 4 | 2] + D_{1354} [ 1 \, 5, 3 \, 4 | 2]+
D_{1543} [ 1 \, 4, 5 \, 3 | 2] \\
&\left.+D_{5234} [ 5 \, 3, 2 \, 4 | 1]+ D_{5324} [ 5 \, 2, 3 \, 4 | 1]  
+ D_{5243} [ 5 \, 4, 2 \, 3 | 1]\right) \; . 
\end{aligned}
\end{equation}
Here $G^{1}_{\{L_i\}}$ is the one-loop correction to the connected correlator defined in \eqref{connecteddef} and $[ij,kl|m]$ is given by
\begin{equation}
[ i \, j, k \, l | m] = d_{i m} \, d_{j m} \, d_{k l} + 
d_{i j} \, d_{k m} \, d_{l m} \period
\end{equation}

We now explain how to reproduce the result above from 
the hexagonalization procedure. 
The general rule of the hexagonalization is that 
for a $n$-point correlation function, one needs
$2(n-2)$ hexagons. Thus for $n$=5 one needs to decompose the surface into
six hexagonal patches. The first step in the calculation is to 
enumerate the Wick contractions at tree level. At tree level, one has the
following twelve graphs:    
\begin{equation}
\begin{aligned}\label{perturbative52s}
G^0_{\{2,2,2,2,2\}} = d_{12} d_{24} d_{43} d_{35} d_{51} + d_{12} d_{23} d_{34} d_{45} d_{51}
+ d_{12}d_{25}d_{53}d_{34}d_{41} \; \, \, \\
+ \, d_{12} d_{23} d_{35}d_{54}d_{41} + 
d_{12} d_{25}d_{54}d_{43}d_{31}+ 
d_{12} d_{24} d_{45} d_{53} d_{31} \; \, \, \\
+ \,  d_{13}d_{34}d_{42}d_{25}d_{51} + d_{13} d_{32} d_{24}d_{45}d_{51}
+ d_{13} d_{35} d_{52} d_{24} d_{41} \; \, \, \\
+ \, d_{13}d_{32}d_{25}d_{54}d_{41}+ 
d_{14} d_{43} d_{32} d_{25} d_{51} +
d_{14}d_{42}d_{23}d_{35}d_{51}  \, . 
\end{aligned}
\end{equation}
\begin{figure}[t]
\begin{center}
\includegraphics[clip,height=8cm]{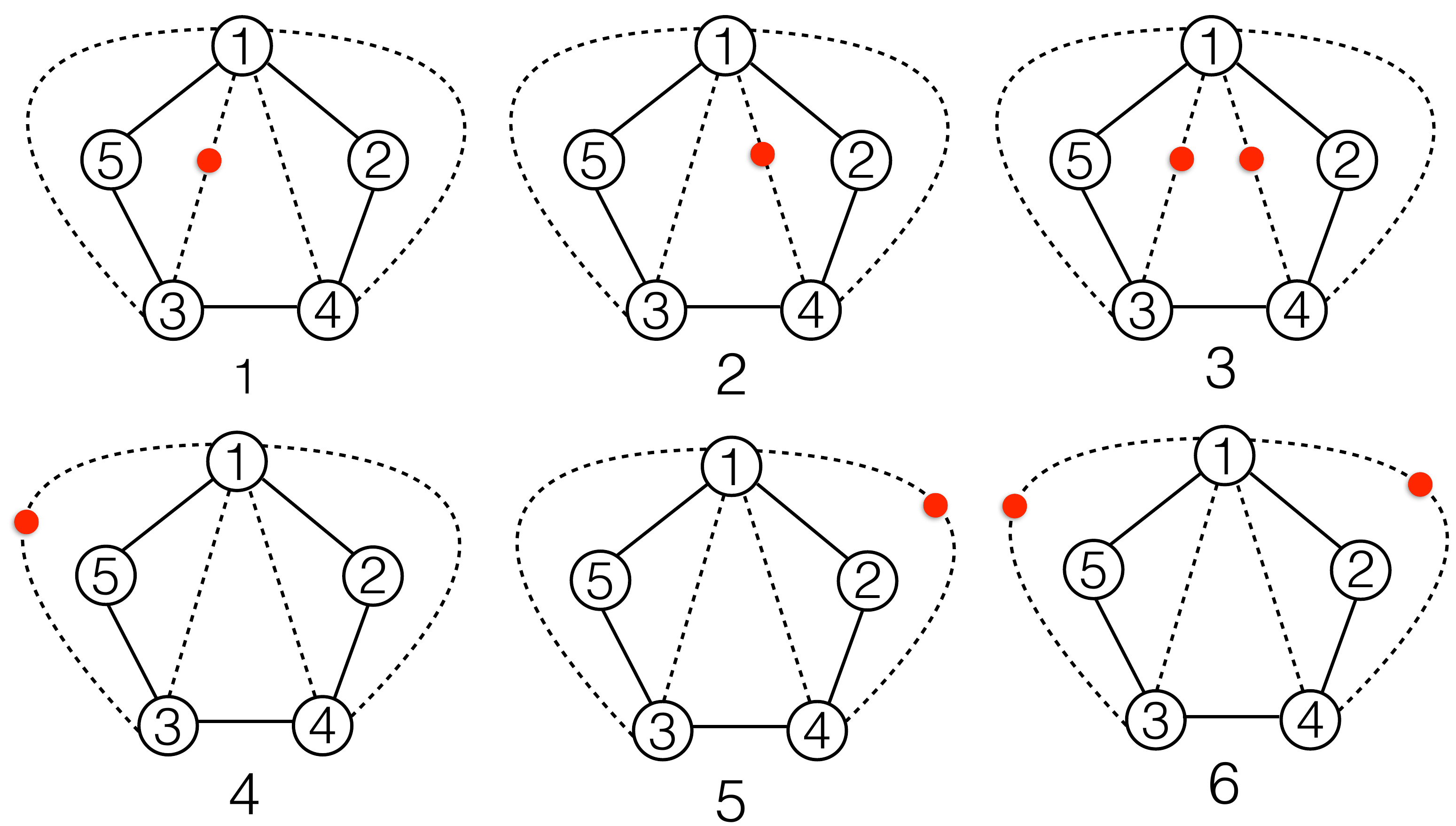} 
\end{center}
\vspace{-0.7cm}
\caption{All the mirror particles contributions for the graph corresponding to the first term in \eqref{perturbative52s}. } 
\label{fig:TheCaseFive202}
\end{figure}

The second step is to decompose each tree-level graph into six hexagons and dress it with mirror particles. Important simplifications at one loop are: 
\begin{enumerate}
\item Only single-particle states on zero-length bridges can contribute.
\item Whenever one introduces more than one particle, they must always be neighbors; namely each of them must share at least one hexagon with some other magnons.
\end{enumerate}
These rules follow simply from the following weak-coupling behavior of the one-particle measure, the mirror energy 
and the hexagon form factor: 
\beq
\begin{aligned}
\mu \sim O(g^2) \, , \quad \quad e^{- \tilde{E}} \sim O(g^2) 
\, , \quad \quad 
\mathfrak{h}(u^{\gamma}, v^{-\gamma}) \sim O(1/g^2) \, . 
\end{aligned}
\eeq
The first two equalities show that 
we can only have single-particle states 
on zero-length bridges. The last equality explains 
why we can have multi-particle contributions: If we 
just take into account the measure 
factor $\mu$, the $n(>1)$ particle contributions 
seem to appear at $O(g^{2n})$. However they get 
enhanced because of the interaction, 
$\mathfrak{h}(u^{\gamma},v^{-\gamma})$.
  
Now, for definiteness, let us 
consider the graph corresponding to the first term in \eqref{perturbative52s}. 
All the one-loop mirror-particle contributions for this case are depicted in figure
\ref{fig:TheCaseFive202}. Summing them up, one gets the result\fn{The way to find the arguments of the functions $\mathcal{M}^{(1)}$ and
$\mathcal{M}^{(2)}$ follows from \eqref{crossratiorule}. 
Here we give further details in how to find them. 
Consider the second diagram of 
figure \ref{fig:TheCaseFive202}. The spacetime 
argument of  $\mathcal{M}^{(1)}$
for this contribution is found by replacing $x_1=0, x_3=1, x_4=\infty$
in the formulas for the cross ratios 
(\ref{eq:TheCrossRatios}) and solving for the 
holomorphic part of $x_2$. 
Similarly, to read the argument for the first diagram of the figure,
one sets $x_1=0, x_5=1, x_3=\infty$ and solves for the holomorphic
part of $x_4$. Finally, the arguments of the function $\mathcal{M}^{(2)}$ 
are the same for the one particle and they are read counterclockwise. 
},
\begin{equation}
\label{eq:oneexampleofContribution}
\begin{aligned}
H_{12435} =& 
 \, 2d_{12}d_{24}d_{43}d_{35}d_{51} \Big[   \mathcal{M}^{(1)}(z, \alpha) +  
 \mathcal{M}^{(1)}(1-w^{-1}, 1- \beta^{-1})  \\ &\quad  \, 
+  \mathcal{M}^{(2)}(z,1-w^{-1},\alpha,1-\beta^{-1}) \Big] \, .
\end{aligned}
\end{equation}
Using \eqref{decagonsum}, one can express it in terms of $m(z)$ as follows\fn{To compare with the perturbative result, it is also useful to express \eqref{Hintermsofm} in terms of the cross ratios $z_{ijkl}$ as follows:
\beq
H_{12435} =2d_{12}d_{24}d_{43}d_{35}d_{51} \left[m(z_{1324})+m(z_{2543})+m(z_{4135})+m(z_{3251})+m(z_{5412})\right]\period
\eeq}:
\beq\label{Hintermsofm}
\begin{aligned}
H_{12435} =&2d_{12}d_{24}d_{43}d_{35}d_{51}\times\\
 &\left[m(z^{-1})+m(1-w^{-1})+m\left(\frac{1-z^{-1}}{1-w^{-1}}\right)+m\left(\frac{z-w}{1-w}\right)+m\left(\frac{z}{w}\right)\right]\period
\end{aligned}
\eeq


The full integrability result is obtained by 
adding the contributions
for all twelve tree-level graphs:
\beq
\begin{aligned}
\left.G^{1}_{\{2,2,2,2,2\}}\right|_{\rm integrability}&=H_{12435}+H_{12345}+H_{12534}+H_{12354}+H_{12543}+H_{12453}\\
&+H_{13425}+H_{13245}+H_{13524}+H_{13254}+H_{14325}+H_{14235}\period
\end{aligned}
\eeq
Adding all the terms and using the properties of $m(z)$ \eqref{propertiesofmz}, we find that the answer perfectly matches the perturbative result \eqref{eq:five20primedata}. 

Note that, although here we chose one particular way of decomposing the five-point function into hexagons, the final result is independent of the way we decompose it. This ``flip invariance'' is
an important consistency check of our results and it will be discussed in section  
 \ref{sec:2particle}.

\subsection{Comparison II: Three $L=2$ and two $L=3$ BPS operators} 

In this section, we compute the correlation function
of three length-two and two length-three 
BPS operators using integrability.
For definiteness, we choose the fourth and the fifth operators to be the length-three operators.
\begin{figure}[t]
\centering
\includegraphics[clip,height=8cm]{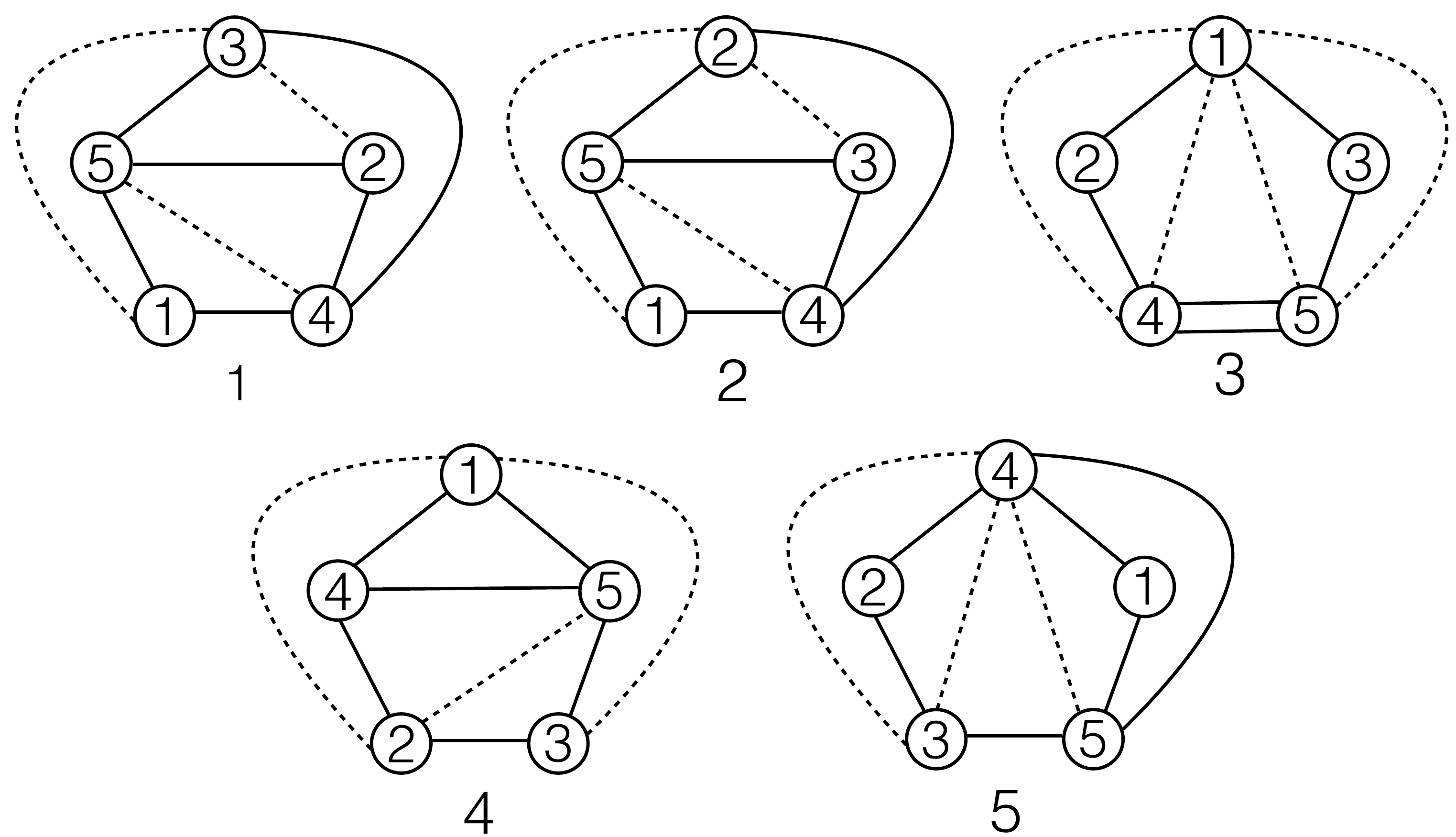} 
\caption{Graphs that contribute to the 
correlation function of three $L=2$ and two $L=3$
BPS operators. The full set of graphs is obtained
by permuting the operators $1, 2$ and $3$ 
in the diagrams $3,4$ and $5$ above. 
The one-loop result is obtained by adding
both the one- and two-particle 
contributions associated with
the zero-length bridges (dashed lines). } 
\label{fig:Five22233}
\end{figure}

 For this correlator, the one-loop perturbative result in \cite{DrukkerPlefka} reads (in our conventions)
\beq
\begin{aligned}\label{eq:TheFive22233}
&\left.G^{1}_{\{2,2,2,3,3\}}(x_1,x_2,x_3,x_4,x_5)\right|_{\rm perturbation}=\\
&\qquad d_{45}G^{1}_{\{2,2,2,2,2\}}(x_1,x_2,x_3,x_4,x_5)+d_{14}d_{15}G^{1}_{\{2,2,2,2\}}(x_2,x_3,x_4,x_5)\\
&\qquad +d_{24}d_{25}G^{1}_{\{2,2,2,2\}}(x_1,x_3,x_4,x_5)+d_{34}d_{35}G^{1}_{\{2,2,2,2\}}(x_1,x_2,x_4,x_5)\comma
\end{aligned}
\eeq
where $G^1_{\{2,2,2,2\}}$ is given by
\beq
G^1_{\{2,2,2,2\}}(x_i,x_j,x_k,x_l)=-D_{ijkl}d_{ik}d_{jl}-D_{ijlk}d_{il}d_{jk}-D_{ikjl}d_{ij}d_{kl}\period
\eeq

\begin{figure}[t]
\centering
\includegraphics[clip,height=4cm]{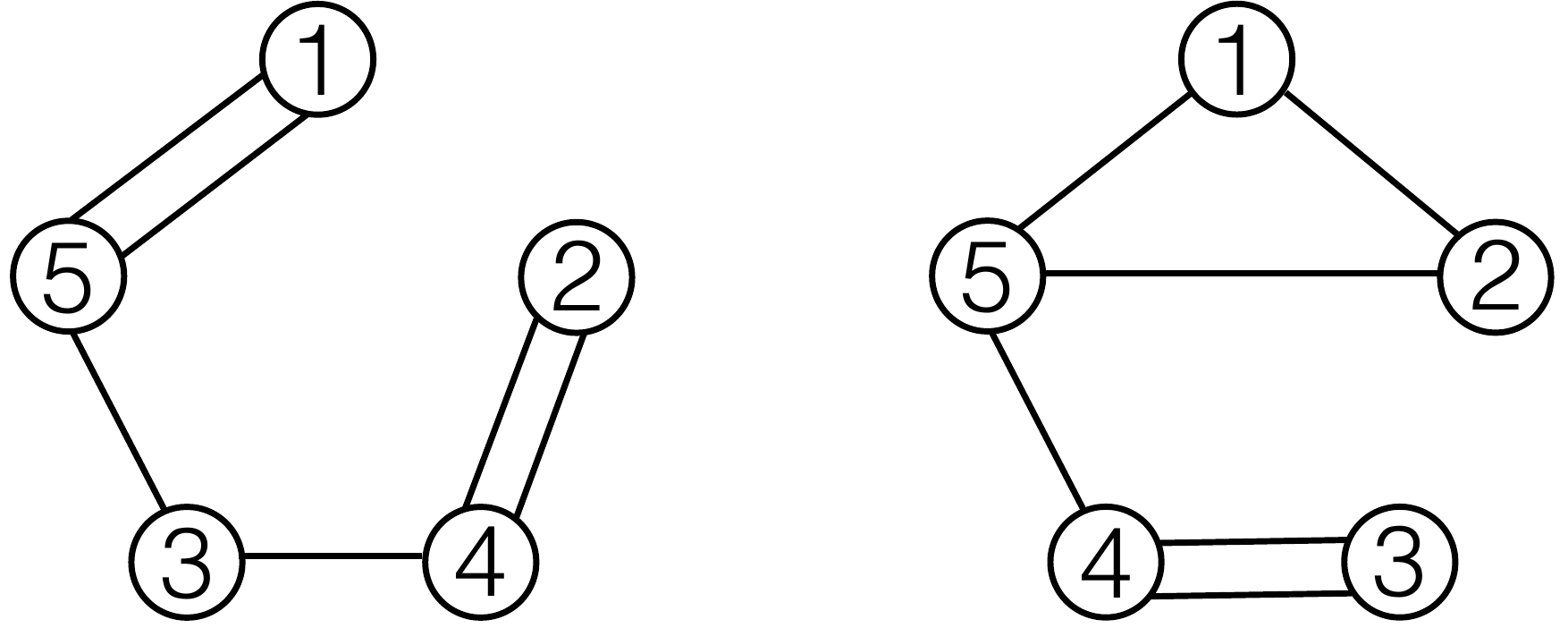} 
\caption{Two examples of non-1EI graphs. 
The 
contributions coming from these types of graphs 
are expected to be zero.} 
\label{fig:FiveNon1EI}
\end{figure}

To reproduce this result from integrability,  one first list 
up the connected tree-level diagrams and then decompose 
them into hexagons. The connected tree-level diagrams 
are divided into two sets, {\it 1-edge irreducible} (1EI) 
graphs and non-1EI graphs \cite{FleuryKomatsu}.  The 
1EI graphs are graphs that are still connected when any
one of its non-zero length bridges are cut.
For the specific correlation function that
we are considering, examples of 1EI graphs
and of non-1EI are given in the figures 
\ref{fig:Five22233} and 
\ref{fig:FiveNon1EI} respectively. In \cite{FleuryKomatsu}, 
we proposed that one only needs to consider 
the corrections coming from 1EI graphs in order to 
compute the correlation functions of the BPS operators. 
In other words, we expect that the mirror-particle 
corrections coming from non-1EI graphs 
add up to zero. 
We have  not been able to show this cancellation explicitly 
for the graphs in figure \ref{fig:FiveNon1EI} since, for that 
purpose, one needs more than two-particle contributions. 
However, we show in section \ref{TheFate1EI} that such 
cancellation indeed takes place for 
next-to-extremal four-point functions. In what follows, we 
compute the one-loop five-point function assuming 
that the non-1EI graphs do not contribute.

The 1EI graphs relevant for the correlation function 
of three length-two and two
length-three BPS operators are shown in figure 
\ref{fig:Five22233}.
The complete set of graphs is composed of the diagrams 1 and 2 and
of all the permutations of the operators 1, 2 and 3 of the diagrams
3, 4 and 5.  At one loop, one has to compute the one-particle
and two-particles contributions associated with the zero-length
bridges. From the diagrams $1$ and $2$, we get 
\begin{equation}
\begin{aligned}
{\rm Diagram } \, 1+ {\rm Diagram } \,2 \,= 
\, d_{41} d_{15} d_{53}d_{34}d_{42}d_{25} 
\, \times \hspace{30mm} \\
\left[  2  \mathcal{M}^{(1)} \left(1-\frac{z}{w}, 1- 
\frac{\alpha}{\beta}\right) + 2  \mathcal{M}^{(1)}
\left(\frac{w-z}{w-1},\frac{\beta-\alpha}{\beta-1}\right) 
+ 2 \, \mathcal{M}^{(1)}\left(1-w^{-1},1-\beta^{-1}\right)\, \right]
\, . 
\end{aligned}
\end{equation}
The diagram 3 is similar to the diagrams that appeared in the previous subsection, see figure \ref{fig:TheCaseFive202}; the only difference is that it has one more propagator:
\beq
\begin{aligned}
{\rm Diagram}\,3 =\, d_{45}H_{13542}\period
\end{aligned}
\eeq
Lastly, the diagrams $4$ and $5$ give
\begin{equation}
\begin{aligned}
{\rm{Diagrams}} \, 4+{\rm{Diagrams}}\, 5 \, = 
\, d_{14} d_{45} d_{51}d_{42}d_{23}d_{35} 
\, \times \hspace{30mm} \\
\left[ \, 3 \, \mathcal{M}^{(1)} \left(\frac{w-1}{z-1},  
\frac{\beta-1}{\alpha-1}\right) +  \, \mathcal{M}^{(1)}
\left(1-z,1-\alpha\right) 
+  \, \mathcal{M}^{(1)}\left(1-w^{-1},1-\beta^{-1}\right)\right.\,  \hspace{10mm} \\
\left.+  \, \mathcal{M}^{(1)}\left( \frac{z(w-1)}{w(z-1)},
\frac{\alpha(\beta-1)}{\beta(\alpha-1)} \right) +   \mathcal{M}^{(2)}
\left(1-z,\frac{z(w-1)}{w(z-1)} \right)  
+  \mathcal{M}^{(2)}
\left(1-w^{-1},\frac{z-1}{w-1} \right)  
\right]
\, .  
\end{aligned}
\end{equation}

Adding all the diagrams (including the permutations),
one can reproduce the perturbative result given in (\ref{eq:TheFive22233}).
 This supports our original assumption that
the sum of the mirror-particle corrections for non-1EI graphs vanishes.

\section{Two-Particle Contributions and 
Flip Invariance \label{sec:2particle}}
We now outline how to compute the two-particle contribution shown in figure \ref{fig:TheTwoMirrorParticlesFig} to get (\ref{eq:twoparticle}). More technical details are explained in Appendices \ref{ap:S-matrix} and \ref{TheIntegrand}.
In addition, we discuss the flip invariance of the 
two-particle result. 
\subsection{The Two-Particle Computation}
To compute the two-particle contribution given in figure \ref{fig:TheTwoMirrorParticlesFig},  one has 
to evaluate the contribution from each hexagon and sum over all the mirror
particle bound states which we insert on the dashed edges.

Let us begin by recalling what the mirror-particle states are. A complete basis of states on the mirror edge is given by multi-particle states made up of various bound states.  Each bound state is labelled by the integer $a$, and the $a$-th mirror bound state $\mathcal{X}$ is made up of a 
pair of ``quarks" $\chi$ and $\dot{\chi}$, each of 
which belongs to the $a$-th anti-symmetric representation of $\mathfrak{su}(2|2)$:
\beq
\label{eq:dotundot}
\begin{aligned}
\chi\comma \,\,\dot{\chi}= \, \, \, &|\psi_{\alpha_1}\cdots\psi_{\alpha_a} \rangle+\cdots \comma \quad &&|\phi_1\psi_{a_1}\cdots \psi_{\alpha_{a-1}}\rangle +\cdots\comma\\
&|\phi_{2}\psi_{\alpha_1}\cdots \psi_{\alpha_{a-1}}\rangle+\cdots.
\quad &&|\phi_1\phi_2\psi_{\alpha_1}\cdots \psi_{\alpha_{a-2}}
\rangle+\cdots.  
\end{aligned}
\eeq
A small complication which arises for the multi-point functions is that the naive basis given above does not reproduce the correct perturbative result. To obtain a match, one needs to dress the basis elements with the so-called $Z$-markers as follows \cite{FleuryKomatsu}:
\beq\label{averageoversign}
\begin{aligned}
&|\psi_{\alpha_1}\cdots\psi_{\alpha_a} \rangle+\cdots \comma \quad &&|Z^{\pm \frac{1}{2}} \phi_1\psi_{a_1}\cdots \psi_{\alpha_{a-1}}\rangle +\cdots\comma\\
&|Z^{\mp \frac{1}{2}} \phi_{2}\psi_{\alpha_1}\cdots \psi_{\alpha_{a-1}}\rangle+\cdots.
\quad &&|\phi_1\phi_2\psi_{\alpha_1}\cdots \psi_{\alpha_{a-2}}
\rangle+\cdots,  
\end{aligned}
\eeq 
The correct result is reproduced after one averages over the two signs. This was shown explicitly for the one-particle contribution in \cite{FleuryKomatsu}. Also for the two-particle contribition, we found that essentially the same prescription (with a little bit of refinement) gives the correct answer. For details of the prescription, see Appendix \ref{ZmarkersSection}.
\begin{figure}[t]
\centering
\includegraphics[clip,height=10.5cm]{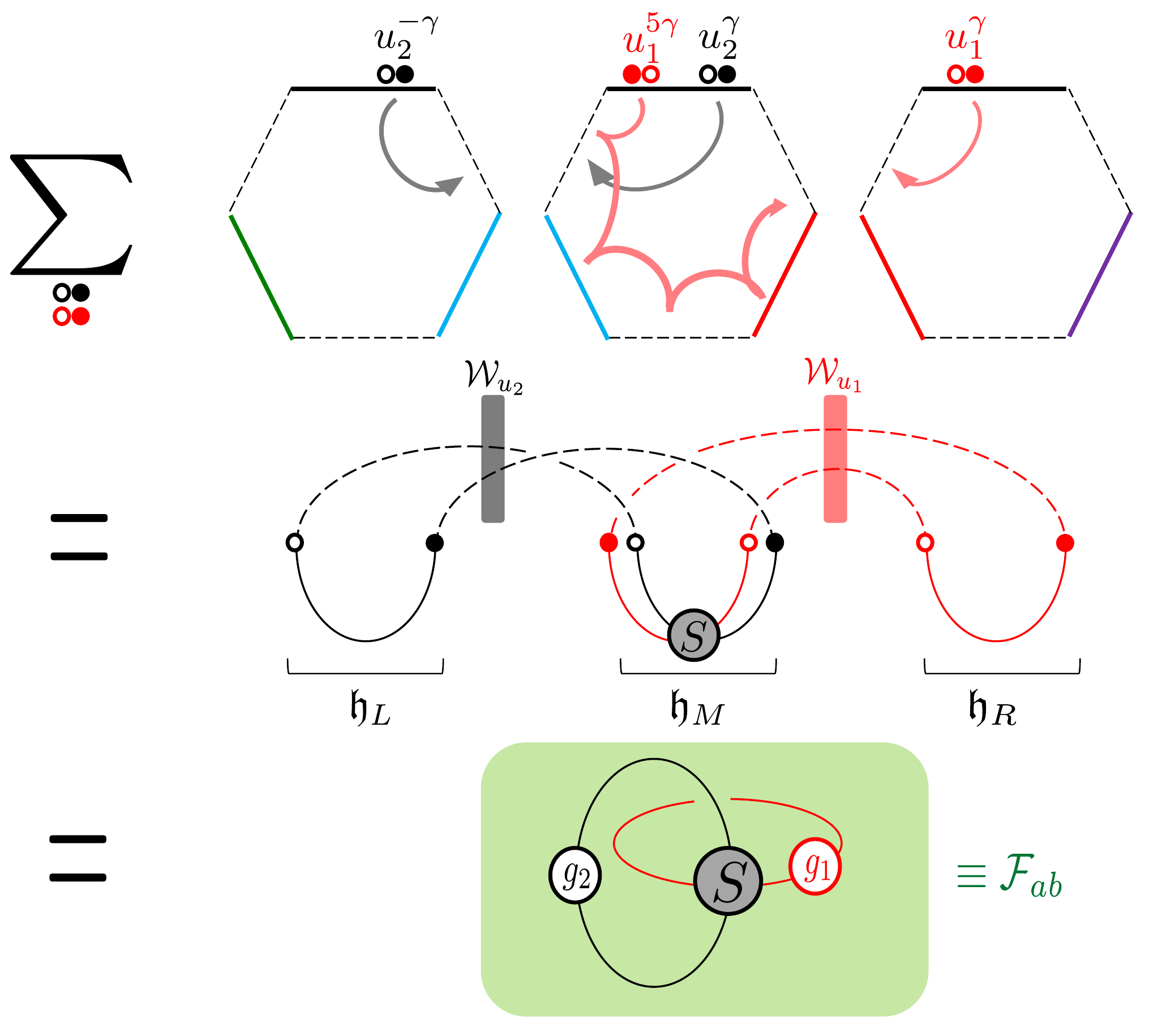} 
\caption{The matrix part for the two-particle contribution. The top line explains how we compute it using the hexagon formalism: We rewrite the middle hexagon by performing the $5\gamma$-mirror transformation to the first particle. The second line shows the pictorial representation of the matrix part: The summation over the flavor indices corresponds to adding dashed curves in the figure, and $\mathcal{W}_{u_i}$'s are the weight factors. The last line gives the final result for the matrix part and the definition of $\mathcal{F}_{ab}$: It is essentially given by two intersecting loops. At the intersection point we insert 
the $\mathfrak{su}(2|2)$ S-matrix (denoted by a gray dot in the figure) and, on each loop, we insert a twist $g_i$. The twist comes from the insertion of the weight factor $\mathcal{W}_{u_i}$ and it produces different phases depending on the flavors. (See also \cite{FleuryKomatsu}).} 
\label{fig:TheCaseFive20}
\end{figure}
%

Having determined the correct basis elements, one can write down\fn{One also needs to average over the choices of the signs in \eqref{averageoversign}, although we did not write it explicitly here.} the two-particle contribution following the general prescription given in \cite{FleuryKomatsu}:
\begin{equation}
\label{eq:ExplainingTwoParticle}
\begin{aligned}
\mathcal{M}^{(2)}(z_1, z_2, \alpha_1,\alpha_2) =
\int \frac{d u_1}{ 2 \pi} \frac{ d u_2}{ 2 \pi} 
 \, \sum_{a =1}^{\infty} \sum_{b =1}^{\infty}\mu_{a}(u_1^{\gamma}) \mu_{b}(u_2^{\gamma}) \sum_{I}\sum_J
\hspace{30mm}  \\
 \mathfrak{h}_L \left[
\bar{\mathcal{X}}_b (u_2^{-\gamma})_J \right]
\, \mathcal{W}\left[\mathcal{X}_b (u_2^{\gamma})_J\right] \, 
\mathfrak{h}_{M}\left[ 
\mathcal{X}_b (u_2^{\gamma})_J \,  
\bar{\mathcal{X}}_a (u_1^{-\gamma})_I\right] \,  
\mathcal{W}\left[\mathcal{X}_a (u_1^{\gamma})_I\right] \, 
\mathfrak{h}_R\left[ \mathcal{X}_a (u_1^{\gamma})_I\right] \, ,  
\end{aligned} 
\end{equation}  
Here $\mathfrak{h}$ denote the hexagon form factors 
and the subscripts $L, M$ and $R$ mean
the left hexagon, the middle hexagon and the right hexagon 
respectively, see 
figure  \ref{fig:TheTwoMirrorParticlesFig}. $\mathcal{W}$'s are the weight factors which incorporate the cross-ratio dependence and $\mu$'s are the measures.
There are two bound state indices $a$ and $b$ 
corresponding to the mirror particles $1$ and $2$.
The indices $I$ and $J$ label the states inside the bound-state module $a$ and $b$ respectively.


The expression \eqref{eq:ExplainingTwoParticle} by itself is not very useful for the actual computation. In what follows, we explain the individual factors and derive a more compact form of the two-particle integrand. Let us first discuss the weight factors $\mathcal{W}$. As discussed in \cite{FleuryKomatsu}, they are determined by the symmetry arguments and consist of the 
a flavor-independent and a flavor-dependent part.
The flavor-independent part is given by
\begin{table}[t]
    \begin{minipage}{0.45\hsize}
    \begin{center}
    \begin{tabular}{r||c|c}
   &${\sf L}$&${\sf R}$\\
    $\psi^{1}$&$+\frac{1}{2}$&$0$\\
    $\psi^{2}$&$-\frac{1}{2}$&$0$\\
    $\phi^{1}$&$0$&$+\frac{1}{2}$\\
    $\phi^{2}$&$0$&$-\frac{1}{2}$
     \end{tabular}
    \end{center}
    \end{minipage}
    \begin{minipage}{0.45\hsize}
    \begin{center}
    \begin{tabular}{r||c|c}
    &${\sf L}$&${\sf R}$\\
    $\psi^{\dot{1}}$&$-\frac{1}{2}$&$0$\\
    $\psi^{\dot{2}}$&$+\frac{1}{2}$&$0$\\
    $\phi^{\dot{1}}$&$0$&$-\frac{1}{2}$\\
    $\phi^{\dot{2}}$&$0$&$+\frac{1}{2}$
    \end{tabular}
    \end{center}
    \end{minipage}
    \caption{The charges of a fundamental magnon under the spacetime and the R-symmetry rotations ${\sf L}$ and ${\sf R}$.\label{tab1}} \label{tab:TheChargesFundamentalMagnonTable}
\end{table}
\begin{equation}
\mathcal{W}_{{\rm{non-flavor}}}\left[\mathcal{X} (u_i^{\gamma})\right] = 
e^{- 2 i \tilde{p}_{u_i} {\rm{log}} \, |z_i|}  \, , 
\end{equation}  
where $\tilde{p}_{u_i}$ is the mirror momentum,
while the flavor-dependent part is given by 
\begin{equation}
\mathcal{W}_{{\rm{flavor}}}\left[\mathcal{X} \right] =    
e^{i J_{\mathcal{X}} \varphi_i} \, e^{i L_{\mathcal{X}} \phi_i} \, e^{i R_{\mathcal{X}} \theta_i}
\, ,
\end{equation}
where $J_{\mathcal{X}}$, $L_{\mathcal{X}}$ and $R_{\mathcal{X}}$ are the eigenvalues of the state $\mathcal{X}$ for the generators $J$, $L$, and $R$ defined below\fn{The fact that only the generators $L$, $R$ and $J$ enter in the weight factor is a consequence of our special kinematics: If the operators are not contained in a single two-dimensional plane, one would need other generators which move the operators away from that two-dimensional plane.
}:
\begin{align}
\begin{aligned}
&J: \quad e^{i J \varphi}Z =e^{i \varphi} Z \comma \qquad e^{i J \varphi}\bar{Z} =e^{-i \varphi} \bar{Z}\comma\\
&L = \frac{1}{2}(L^{1}_{\; 1}-L^{2}_{\; 2} - L^{\dot{1}}_{\; \dot{1}}
+L^{\dot{2}}_{\; \dot{2}}) \, , \quad \quad 
R = \frac{1}{2}(R^{1}_{\; 1}-R^{2}_{\; 2} - R^{\dot{1}}_{\; \dot{1}}
+R^{\dot{2}}_{\; \dot{2}}) \, , 
\end{aligned}
\end{align}
with the angles 
\beq
e^{i\phi_i}=\sqrt{\frac{z_i}{\bar{z}_i}}\comma\quad e^{i\theta_i}=\sqrt{\frac{\alpha_i}{\bar{\alpha}_i}}\comma \qquad e^{i\varphi_i}=\sqrt{\frac{\alpha_i\bar{\alpha}_i}
{z_i\bar{z}_i}}\period
\eeq
The eigenvalues $L_{\mathcal{X}}$ and $R_{\mathcal{X}}$ can be read off from the charges of the fundamental magnons listed in table  
\ref{tab:TheChargesFundamentalMagnonTable}. 

It then remains to evaluate the hexagon form factors $\mathfrak{h}$'s. The hexagon form factors consist of the dynamical part, which is an overall scalar factor, and the matrix part, which depends on the flavor. To evaluate each factor, it is convenient to perform the mirror transformations to the middle hexagon and rewrite it as
\begin{equation}
\mathfrak{h}_{M} \left[
\mathcal{X}_b (u_2^{\gamma})_{J}    
\, \bar{\mathcal{X}}_a (u_1^{-\gamma})_I \right] =
(-1)^{a} \, \mathfrak{h}_{M} \left[ 
\mathcal{X}_a (u_1^{5 \gamma})_{I} \,     
\mathcal{X}_b (u_2^{\gamma})_J \right] \, . 
\end{equation}   
The sign $(-1)^a$ and the replacement 
$\bar{\mathcal{X}}$ to $\mathcal{X}$
for the entry with $u_1$ 
are the consequences of the crossing 
rules given in \cite{BKV}. One can then split $\mathfrak{h}_{M} [ \,
\mathcal{X}_a (u_1^{5 \gamma})_{I} \,     
\mathcal{X}_b (u_2^{\gamma})_J ]$ into the dynamical part $h$ and the matrix part ${\rm MP}$ as follows:
\beq
\mathfrak{h}_{M}\left[  
\mathcal{X}_a(u_1^{5 \gamma} )_I 
\mathcal{X}_b(u_2^{\gamma})_J  \right] = (-1)^{F_1 F_2} \, 
h_{ab} (u_1^{5 \gamma}, u_2^{ \gamma} ) \, 
{\rm MP}_{ab,IJ}(u_1^{\gamma}, u_2^{\gamma}) \, \period
\eeq
Here we used the invariance of the matrix part under the $4\gamma$ tranformation. The dynamical part can be evaluated by using the property
\beq
h_{ab} (u_1^{5 \gamma}, u_2^{ \gamma} )=\frac{1}{h_{ba}(u_2^{ \gamma}, u_1^{ \gamma} )}\comma
\eeq
and the explicit weak-coupling expansions given in Appendix \ref{WeakCouplingExpansions}. On the other hand, the matrix part ${\rm MP}$ is given essentially by the matrix elements of the $\mathfrak{su}(2|2)$ S-matrix. For a pictorial explanation, see figure \ref{fig:TheCaseFive20}.

The hexagon form factors for the left and the right hexagons can also be represented pictorially\fn{For these hexagon form factors, the dynamical factors are just unity.} as shown in figure \ref{fig:TheCaseFive20}. Using such pictorial representations, it is easy to see that the flavor-dependent part 
\beq
\mathcal{F}_{ab}\equiv \sum_{I,J}\mathfrak{h}_L \left[
\bar{\mathcal{X}}_{b,J} \right]
\, \mathcal{W}_{\rm flavor}\left[\mathcal{X}_{b,J}\right] \, 
 {\rm MP}_{ab,IJ}(u_1^{\gamma}, u_2^{\gamma}) \,  
\mathcal{W}\left[\mathcal{X}_{a,I}\right] \, 
\mathfrak{h}_R\left[ \mathcal{X}_{a,I}\right] \,
\eeq
becomes the quantity depicted in the last line of figure \ref{fig:TheCaseFive20}.
As shown there, it is essentially given by the matrix elements of the $\mathfrak{su}(2|2)$ $S$-matrix dressed by the weight factors. 

The computation of $\mathcal{F}_{ab}$ is the most complicated task needed for this work. One first needs to construct the bound-state $S$-matrices, then multiply them with the weight factors and compute traces. The computation simplifies slightly owing to the restricted kinematics we chose: All the generators $L$, $R$ and $J$ which appear in the weight factor act diagonally on the bound-state basis. Therefore, when we perform the computation, one only needs the {\it diagonal} components of the S-matrix,
\begin{equation}
S \, \cdot \, | u_1^{\gamma}, a \rangle_{I} \otimes    
 | u_2^{\gamma}, b \rangle_{J}  \, \rightarrow \, 
(S_{ab})_{I \, J}^{J \, I}
\;  | u_2^{\gamma}, b \rangle_{J} \otimes    
 | u_1^{\gamma}, a \rangle_{I} \, . 
\end{equation}
This feature makes the computation slightly easier although it is still a tedious task. See Appendix \ref{BoundStateSection} for the detail of the computation. Using the result there, one can then express the two-particle integrand as follows:
\beq\label{M2integrandmaintext}
\begin{aligned}
\mathcal{M}^{(2)}(z_1, z_2, \alpha_1,\alpha_2) =
\int \frac{d u_1}{ 2 \pi} \frac{ d u_2}{ 2 \pi} 
 \, \sum_{a =1}^{\infty} \sum_{b =1}^{\infty}\frac{\mu_{a}(u_1^{\gamma}) \mu_{b}(u_2^{\gamma})}{h_{ba}(u_2^{\gamma},u_1^{\gamma})} e^{-i \tilde{p}_{a}(u_1)\log |z_1|}e^{-i \tilde{p}_{b}(u_2)\log |z_2|}\mathcal{F}_{ab}\period
\end{aligned}
\eeq
A more explicit form of the integrand is given in Appendix 
\ref{TheIntegrand} while the weak-coupling expansions of 
various quantities are listed in Appendix \ref{WeakCouplingExpansions}. By performing the integration in \eqref{M2integrandmaintext}, one arrives at the expression \eqref{eq:twoparticle}.

\subsection{Flip invariance\label{subsec:flipinvariance}}

In this subsection, we will discuss the flip invariance of the 
decagon, i.e. we are going to show that the contribution 
of the mirror particles is independent of the way 
one decomposes the decagon into hexagons. The decagon 
which is a polygon with both five physical and five mirror 
edges appears
for example in the computation of the
five-point function of length two BPS operators. 
As discussed before, we are considering  
a restricted kinematics where all five operators
are in a plane (we also impose an analogous constraint on the R-charge
polarizations). The configuration is characterized 
by the set of cross ratios given in (\ref{eq:TheCrossRatios}). 
\begin{figure}[t]
\centering
\includegraphics[clip,height=7cm]{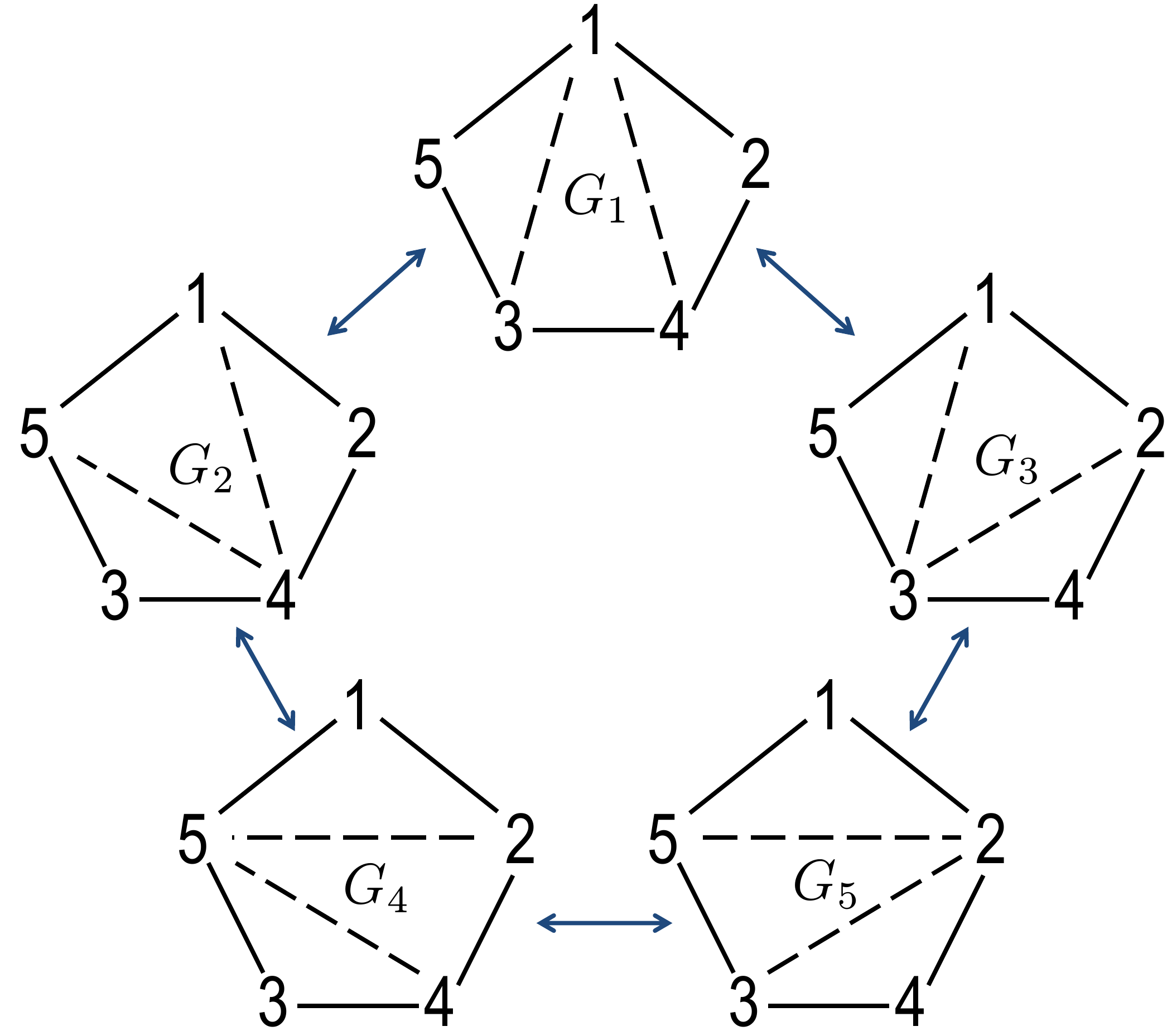} 
\caption{The flip invariance of the decagon. Each diagram in the figure corresponds to a different way of cutting the decagon into hexagons. They all give the same answer.} 
\label{fig:cutting}
\end{figure}

At one loop level, we have two one-particle contributions 
coming from each zero-length bridge and a 
two-particle contribution, which represents the 
interaction between two one-particle states (see figure \ref{fig:TheTwoMirrorParticlesFig}).
Depending on how we cut the decagon into three hexagons, we obtain 
the following expressions\fn{Here, for simplicity, we did not write the $R$-charge cross-ratio argument of
$\mathcal{M}^{(1)}$ and $\mathcal{M}^{(2)}$. 
}: 
\beq
\begin{aligned}
G_1&=\mathcal{M}^{(1)}\left(z\right)+\mathcal{M}^{(1)}\left(1-w^{-1}\right)
+\mathcal{M}^{(2)}
\left(z,1-w^{-1}\right)\comma\\
G_2&=\mathcal{M}^{(1)}
\left(\frac{1}{1-w^{-1}}\right)+\mathcal{M}^{(1)}\left(\frac{z}{w}\right)+
\mathcal{M}^{(2)}\left(\frac{1}{1-w^{-1}},\frac{z}{w}\right)\comma\\
G_3&=\mathcal{M}^{(1)}
\left(\frac{1-w^{-1}}{1-z^{-1}}\right)+\mathcal{M}^{(1)}\left(\frac{1}{z}\right)
+\mathcal{M}^{(2)}\left(
\frac{1-w^{-1}}{1-z^{-1}},\frac{1}{z}\right)\comma
\\
G_4&=\mathcal{M}^{(1)}
\left(\frac{w}{z}\right)+\mathcal{M}^{(1)}\left(\frac{w-z}{w-1}\right)+
\mathcal{M}^{(2)}\left(\frac{w}{z},\frac{w-z}{w-1}\right)\comma\\
G_5&=\mathcal{M}^{(1)}
\left(\frac{w-1}{w-z}\right)+\mathcal{M}^{(1)}\left(\frac{1-z^{-1}}{1-w^{-1}}\right)
+\mathcal{M}^{(2)}\left(
\frac{w-1}{w-z},\frac{1-z^{-1}}{1-w^{-1}}\right)\period\\
\end{aligned}
\eeq
Here $G_i$ denotes 
the expression coming from each configuration in figure \ref{fig:cutting}.
Using the expressions for $\mathcal{M}^{(1)}$ and $\mathcal{M}^{(2)}$
given in (\ref{eq:oneparticle}) and (\ref{eq:twoparticle}),
one can show that all the $G_i$ above 
are equal. This establishes
the flip invariance of the decagon.

\section{Next-to-Extremal Four-Point Functions} \label{TheFate1EI}
In \cite{FleuryKomatsu}, it was conjectured that one only needs to compute the mirror-particle corrections associated with {\it 1-edge irreducible} (1EI) graphs in order to reproduce the correlation functions of BPS operators. This is equivalent to saying that the mirror-particle corrections associated with a non-1EI graph cancel among themselves\fn{
Recently in \cite{EdenNew}, an interesting proposal was made regarding the non-1EI graphs. They proposed that one should multiply the color factors to each hexagons and attributed the cancellation of the non-1EI graphs to such color factors. Nicely as we show here, the color factors do not seem necessary to observe the cancellation. This and related issues will be discussed more in detail in \cite{nonplanarpaper,nonplanarpaper2}.
}. 
Due to the fact that non-1EI graphs have multiple length-zero
bridges, showing 
the cancellation requires the knowledge of 
the multi-particle contribution. In this section, using the two-particle contribution computed in the previous section, we show this cancellation explicitly for the so-called {\it next-to-extremal four-point functions} at one loop. To show it for more general one-loop four-point functions, one needs three-particle contributions and we will leave it for future investigations.


Let us consider the four-point function
of three length-two BPS operators and one length-four BPS operator 
which we choose to be the operator four.
This is an example of a next-to-extremal correlator because
the lengths satisfy the defining condition $L_4=L_1+L_2+L_3-2$.  
This kind of correlators is known to be protected \cite{Extremal4,Extremal2,Extremal3}. Namely the quantum corrections to this correlator must vanish.
\begin{figure}[t]
\centering
\includegraphics[clip,height=4cm]{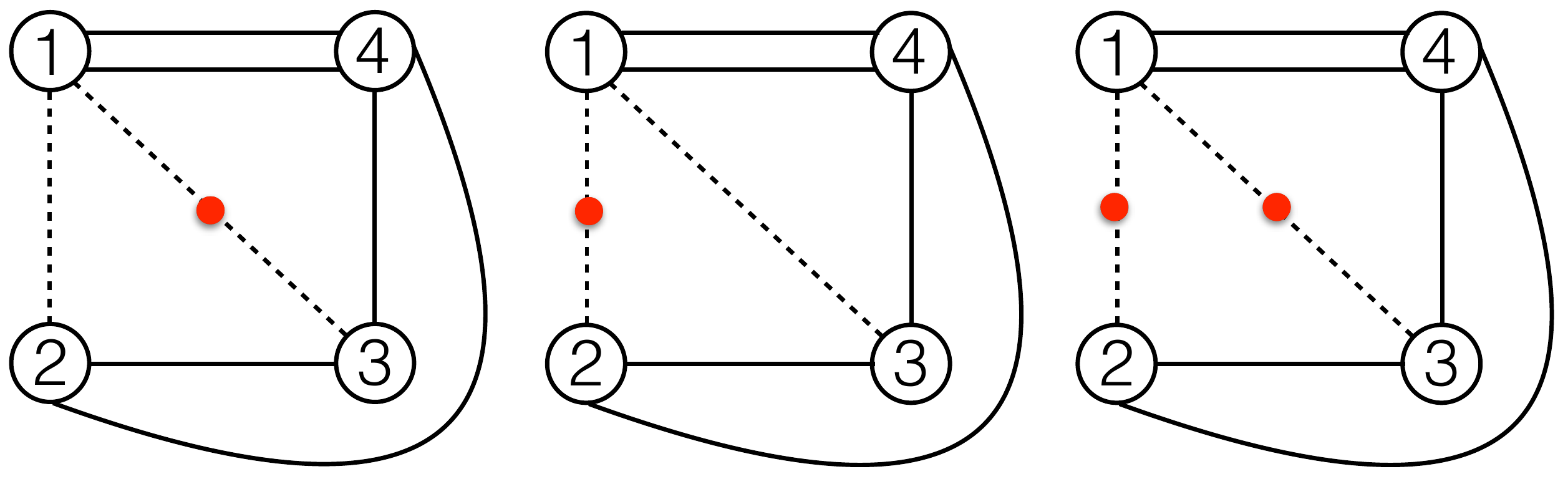} 
\caption{An example of the tree-level Wick contraction 
for $G_{\{2,2,2,4\}}$ and its mirror-particle corrections.
Other diagrams, which are also non-1EI, can be 
obtained by permuting the indices of the length-$2$ operators. 
The sum of the contributions depicted above vanishes 
in agreement with the non-renormalization property of 
the next-to-extremal correlator.
} 
\label{fig:The2224}
\end{figure}

From the hexagonalization point of view, the computation of this correlator only
involves non-1EI graphs. An example of non-1EI graphs and its mirror-particle corrections at one loop are given in figure \ref{fig:The2224}. The remaining ones can be obtained by permuting the indices of the external operators.
Using the result in the previous section, one can compute it explicitly as
\begin{equation}
\begin{aligned}
{\rm{Figure}} \, \, \ref{fig:The2224} \; = \; 
d_{14}^2 d_{23}d_{24}d_{34} \, \left[  \, \mathcal{M}^{(1)}(1-z^{-1},1-\alpha^{-1})+
\mathcal{M}^{(1)}\left(\frac{1}{1-z},\frac{1}{1-\alpha}\right)\right. \hspace{10mm} \\
\left.+ \, \mathcal{M}^{(2)}\left(1-z^{-1}, \frac{1}{1-z}, 1-\alpha^{-1},\frac{1}{1-\alpha}\right)
 \right] \, .
\end{aligned}
\end{equation}
Using \eqref{decagonsum} and the properties of $m(z)$ given in \eqref{propertiesofmz}, one obtains
\beq
{\rm{Figure}} \, \, \ref{fig:The2224} \; =d_{14}^2 d_{23}d_{24}d_{34}\left[m\left(\frac{z}{z-1}\right)+m\left(\frac{1}{1-z}\right)+2 m(1)+m(0)\right]=0\period
\eeq
This proves the absence of one-loop corrections to this four-point function.

\begin{figure}
\centering
\begin{minipage}{0.45\hsize}
\includegraphics[clip,height=4cm]{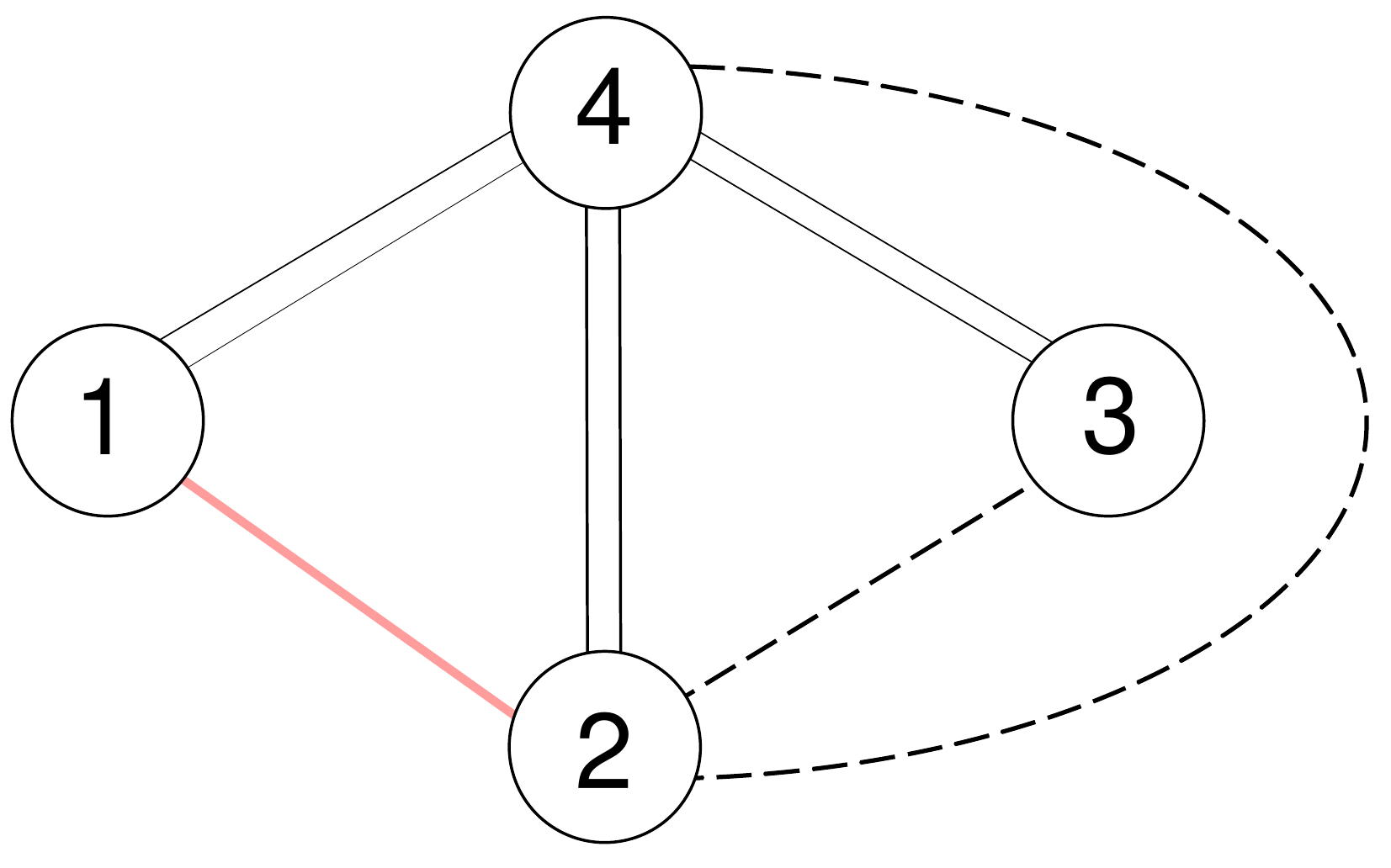}
\end{minipage}
\begin{minipage}{0.45\hsize}
\includegraphics[clip,height=4cm]{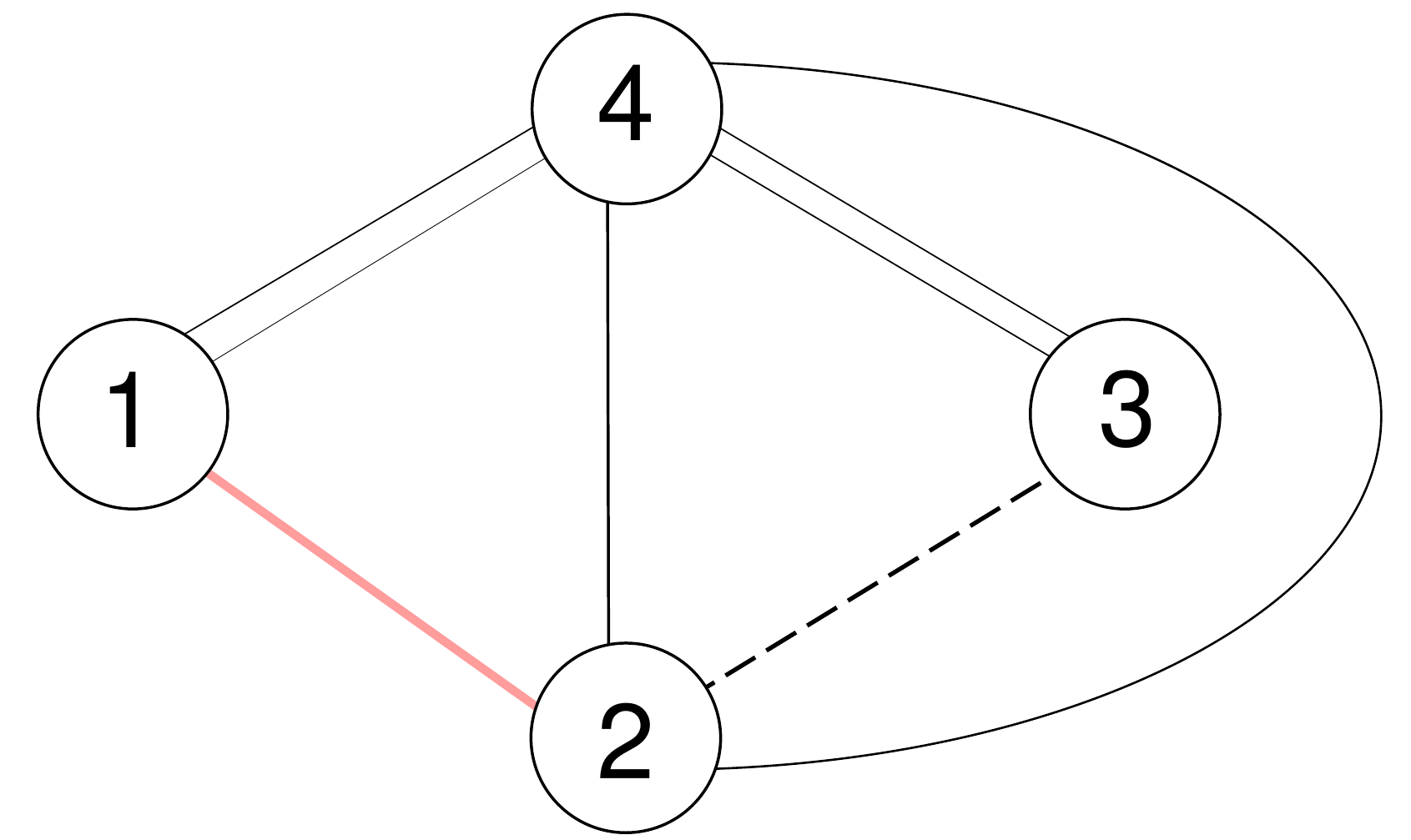}
\end{minipage}
\caption{Examples of graphs for the extremal four-point functions. The solid black lines are the propagators while the red thick lines are the additional propagator one adds to make them into graphs for the next-to-extremal four-point functions. In the left figure, the resulting next-to-extremal graph contains two length-zero bridges (denoted by the dashed lines), while in the right figure, it only has one length-zero bridge whose cross ratio is unity.}
\label{fig:exttonext}
\end{figure}
Although we have focused on one particular example so far, the argument presented here can be readily applied to general next-to-extremal four-point functions. To see this, one just needs to use the fact that the graphs for the next-to-extremal four-point functions (satisfying $L_4=L_1+L_2+L_3-2$) can be obtained by
\begin{enumerate}
\item First drawing the graphs for the extremal correlator $G_{L_1-1,L_2-1,L_3,L_4}$.
\item Then adding one propagator between $\mathcal{O}_1$ and $\mathcal{O}_2$.
\end{enumerate} 
It is straightforward to see that the graphs for the extremal correlator contain three or two length-zero bridges. Thus, if we add one propagator as in step 2 above, we are left with graphs with two zero-length bridges or graphs with a single zero-length bridge (see figure \ref{fig:exttonext}). The graphs with two zero-length bridges have the same topology as the one in figure \ref{fig:The2224}, and therefore the contributions from such a graph add up to zero. On the other hand, for the graphs with a single zero-length bridge, the cross ratios associated with the zero-length bridge are $1$. By taking the limit $z,\bar{z}\to 1$ and $\alpha,\bar{\alpha}\to 1$ in the one-particle mirror contribution \eqref{eq:oneparticle}, one can show that the mirror-particle corrections to these graphs also vanish.  
Thus, in summary, the one-loop correction to the next-to-extremal four-point functions always vanishes, in agreement with the non-renormalization theorem shown in \cite{Extremal3}. 

Note also that the cancellation proven in this section holds at the level of individual graphs. Therefore, one can use the result here to show that many of the non-1EI graphs (that appear in more general four-point functions) do not contribute at one loop. The only graphs that are not covered by the discussion here are the graphs which contain three length-zero bridges, such as the ones that appear in the extremal four-point functions. It would be an interesting future problem to compute the three-particle contributions and show the cancellation in full generality.

\section{Summary and Possible Applications} \label{sec:conclusion}

In this paper, we have computed the 
two-particle mirror contribution at one loop. 
Using its result, we computed two
five-point correlators using integrability and 
have found agreement with the
perturbative data.  In addition, we proved that the contributions from non-1EI graphs that appear in the next-to-extremal four-point functions add up to zero, thereby giving supports to the prescription in the previous paper \cite{FleuryKomatsu} which states that one only needs to consider corrections from 1EI graphs. 

In this paper, we only considered the five-point functions in a restricted kinematics where all the operators lie in a two-dimensional plane. It would be an interesting future direction to generalize the computation performed here to the five-point function in general kinematics. As already mentioned, to study the general kinematics, one needs to include in the weight factors the generators that take the operators away from the plane. As a consequence, the non-diagonal elements of the mirror $S$-matrix (computed in Appendix \ref{BoundStateSection}) will also show up in the computation. Although this makes the computation slightly more involved, there is no additional conceptual difficulty in doing this.


The techniques developed in this paper, in particular
the $Z$-marker dressing and the calculation of the mirror
bound state $S$-matrix can be used to evaluate the higher-particle contributions which are necessary for computing the higher-loop corrections and the non-planar correlators. In particular, it is interesting to compute the two-(and three-)loop four-point functions and see if one can reproduce the perturbative data \cite{AllThreeLoop}.

Another interesting application would be to use the hexagonalization to evaluate complicated Feynman integrals. Recently in \cite{BassoDixon}, they succeeded in evaluating particular sets of fishnet diagrams, which generalize the conformal ladder integrals, by using the hexagonalization (and also by the pentagon OPE). It would be interesting to try to extract other integrals that have not yet been computed, such as the double box integrals, by using the two- and higher-particle contributions. For this purpose, it would perhaps be useful to apply the hexagonalization approach to the strongly deformed planar $\mathcal{N}=4$ SYM proposed recently \cite{KazakovI, SiegDeformed, CaetanoKazakov,FishnetI,FishnetII,FishnetIII} since many quantities in this theory admit only a single Feynman diagram at each loop order, and it should be easier to identify the contribution from a given Feynman graph.
%
%

\section*{Acknowledgements} 

We acknowledge discussions with and 
comments by B.~Basso, T.~Bargheer, J.~Caetano, 
V.~Goncalves, V.~Kazakov, G.~Korchemsky. Especially, we thank
P.~Vieira for helping us with the evaluation 
of the integral, in particular by suggesting the basis of functions
used. 
T. F. would like to thank the warm hospitality of 
the Perimeter Institute and of the 
Laboratoire de Physique Th\'eorique
de l'Ecole Normale Sup\'erieure were a large part of this work was done.
T. F. would like to thank CAPES/INCTMAT process
88887.143256/2017-00, the Perimeter Institute 
and the European Research Council (Programme ERC-2012- AdG 320769 AdS-CFT-solvable) for financial support. 
S.K. is supported by the US Department of Energy.

\appendix

\section{The $Z$-marker Dressing} \label{ZmarkersSection}

As discussed in \cite{FleuryKomatsu}, 
the naive basis for the mirror bound states 
does not reproduce the perturbative data and 
it is necessary to dress it with $Z$-markers. 
In this section, we explain how to dress the two-particle 
states by $Z$-markers\fn{At the moment, we do not have
a clear-cut physical explanation for the prescription.
However, adding the $Z$-markers removes all the unwanted square-root cuts of the integrand and makes 
all the states of the bound-state multiplet 
produce a non-zero contribution to the integral. This means that it should be somehow related to the supersymmetry. 
}.

\subsection{Review of $Z$-marker dressing for one-particle states}
Let us first briefly review the rule to dress the one-particle states. 
The $a$-th bound state $\mathcal{X}$ is made up of a 
pair of ``quarks" $\chi$ and $\dot{\chi}$, each of 
which belongs to the $a$-th anti-symmetric representation of $\mathfrak{psu}(2|2)$:
\beq
\begin{aligned}
\chi\comma \,\,\dot{\chi}= \, \, \, &|\psi_{\alpha_1}\cdots\psi_{\alpha_a} \rangle+\cdots \comma \quad &&|\phi_1\psi_{a_1}\cdots \psi_{\alpha_{a-1}}\rangle +\cdots\comma\\
&|\phi_{2}\psi_{\alpha_1}\cdots \psi_{\alpha_{a-1}}\rangle+\cdots\comma 
\quad &&|\phi_1\phi_2\psi_{\alpha_1}\cdots \psi_{\alpha_{a-2}}
\rangle+\cdots\comma 
\end{aligned}
\eeq

To correctly reproduce the four-point function
perturbative data, 
one must dress the states containing 
bosons $\phi_{1}$ and $\phi_{2}$ (and their dotted 
counter parts) as follows:
\beq
\begin{aligned}
\text{$+$ dressing}: \; \; &\phi_1 \to Z^{\frac{1}{2}} \, 
\phi_1 \comma \quad 
\phi_2 \to Z^{-\frac{1}{2}} \, \phi_{2}\comma\quad 
\dot{\phi}_1\to Z^{-\frac{1}{2}} \, 
\dot{\phi}_1\comma \quad \dot{\phi}_2\to Z^{\frac{1}{2}}
\, \dot{\phi}_2\comma\\
\text{$-$ dressing}:\; \; &\phi_1 \to Z^{-\frac{1}{2}}
\, \phi_1 \comma 
\quad \phi_2 \to Z^{ \frac{1}{2}} \, \phi_{2}\comma\quad \dot{\phi}_1
\to Z^{\frac{1}{2}} \, \dot{\phi}_1\comma \quad \dot{\phi}_2\to 
Z^{- \frac{1}{2}}
\dot{\phi}_2\period
\end{aligned}
\eeq
and average over $+$ and $-$ dressings at the end of the computation. 
(Note that the dressings for the left $\mathfrak{psu}(2|2)$ and 
the right $\mathfrak{psu}(2|2)$ are different). 
For instance, for the fundamental magnons, we dress 
the states in the following way:
\beq\label{howeachpm}
\begin{aligned}
&\text{$+$ dressing}:\\
&\quad D_{\alpha\dot{\alpha}}\to 
D_{\alpha\dot{\alpha}}\comma\quad 
\Phi_{1 2}\to Z \, \Phi_{1 2}\comma\quad 
\Phi_{2 1}\to Z^{-1} \, \Phi_{2 1}
\comma\quad \Phi_{1 1}\to 
\Phi_{1 1}\comma\quad \Phi_{2 2}\to \Phi_{2 2}\comma\\
&\quad \Psi_{1\dot{\alpha}}\to Z^{ \frac{1}{2}} \, 
\Psi_{1\dot{\alpha}}\comma \quad \Psi_{2\dot{\alpha}}\to 
Z^{- \frac{1}{2}} \, 
\Psi_{2\dot{\alpha}}\comma\quad \Psi_{\alpha \dot{1}}
\to Z^{- \frac{1}{2}} \, \Psi_{\alpha \dot{1}}
\comma \quad \Psi_{\alpha \dot{2}}\to Z^{
\frac{1}{2}} \, \Psi_{\alpha \dot{2}}\period\\
&\text{$-$ dressing}:\\
&\quad D_{\alpha\dot{\alpha}}\to 
D_{\alpha\dot{\alpha}}\comma\quad \Phi_{12}\to Z^{-1}
\, \Phi_{12}\comma\quad \Phi_{21}\to Z \, \Phi_{21}
\comma\quad \Phi_{11}\to \Phi_{11}\comma\quad \Phi_{22}\to \Phi_{22}
\comma\\
&\quad \Psi_{1\dot{\alpha}}\to Z^{-\frac{1}{2}}
\, \Psi_{1\dot{\alpha}}\comma \quad \Psi_{2\dot{\alpha}}\to 
Z^{\frac{1}{2}} \, \Psi_{2\dot{\alpha}}\comma\quad 
\Psi_{\alpha 1}\to Z^{\frac{1}{2}} \, \Psi_{\alpha 1}\comma \quad 
\Psi_{\alpha 2}\to Z^{-\frac{1}{2}} \, \Psi_{\alpha 2}\period
\end{aligned}
\eeq
\begin{figure}[t]
\centering
\includegraphics[clip,height=4.5cm]{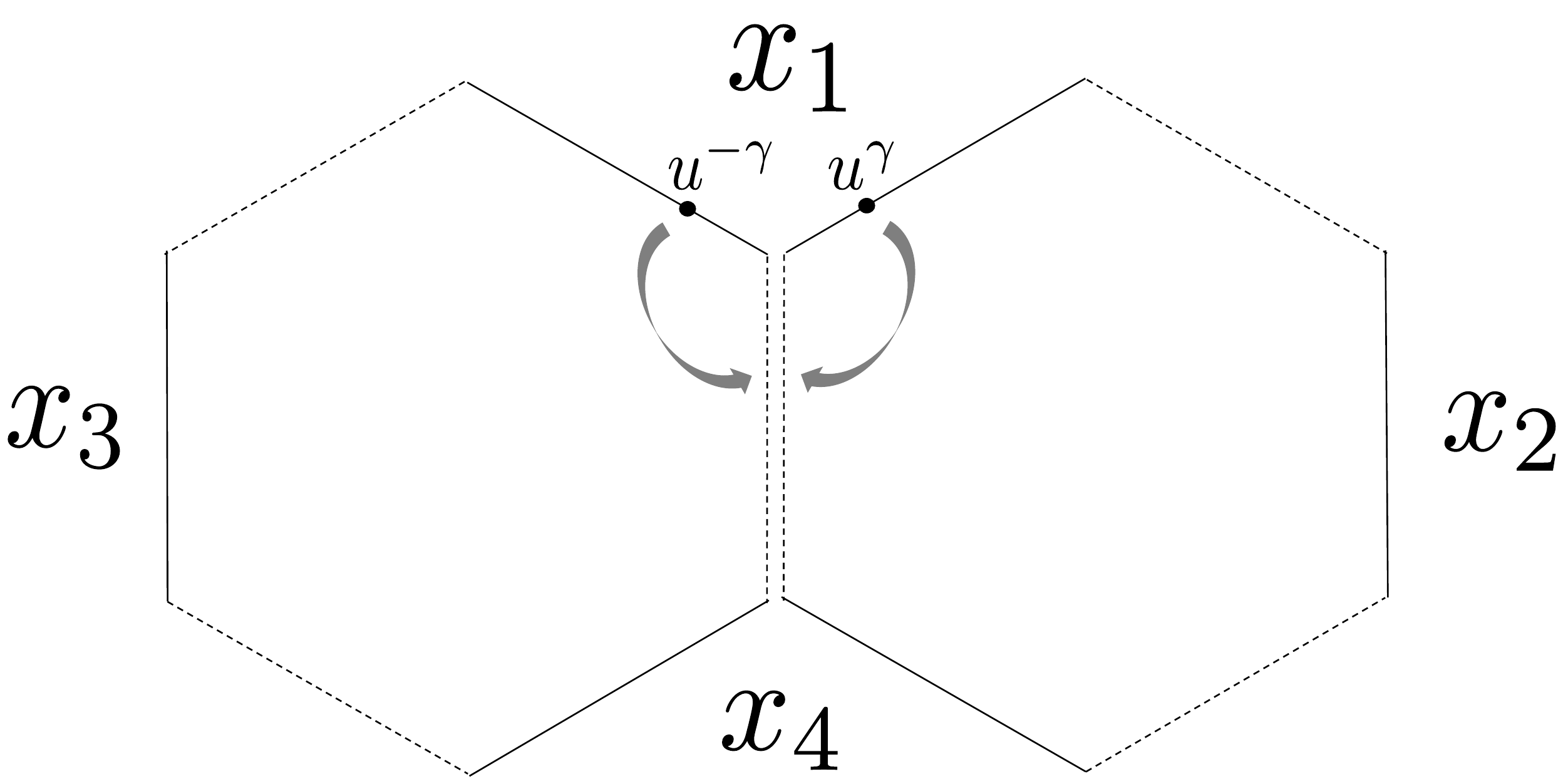}
\caption{The configuration of mirror particles which produces the one-particle contribution. One can glue the edge $14$ by either putting a virtual-particle pair on the edge $1$.\label{fig:1pt}}
\end{figure}

Let us also recall how the weight factors are affected by the Z-marker dressings.
In what follows, we only discuss  
the fundamental magnon, but the generalization to the bound states is straightforward. 
To figure out the effect of the dressing, it is useful to think of the gluing as the process of putting virtual 
particle pairs in the adjacent physical edges and making
the edges entangled \cite{FleuryKomatsu}.  (See also figure \ref{fig:1pt}). Let us consider the 
configuration depicted in figure \ref{fig:1pt}, and try to 
glue the mirror edge $14$ by putting a virtual particle 
pair on the edge $1$. In this case, the 
particle pairs are dressed in the following way:
\beq
\begin{aligned}
\underbrace{|Z^{-t_{\mathcal{X}}^{(\pm)}}\bar{\mathcal{X}}(u^{-\gamma})\rangle}_{\text{left hexagon}} \otimes \underbrace{|\mathcal{X}(u^{\gamma})Z^{+t_{\mathcal{X}}^{(\pm)}}\rangle}_{\text{right hexagon}}\period
\end{aligned}
\eeq
Here the superscript $(\pm)$ in $t_{\mathcal{X}}^{(\pm)}$ denotes the choice of the $+$ or the $-$ dressing and $t_{\mathcal{X}}^{(+)}(=-t_{\mathcal{X}}^{(-)})$ is determined from \eqref{howeachpm} as
\beq\label{tchilist}
\begin{aligned}
&\quad t_{D_{\alpha\dot{\alpha}}}^{(+)}=0\comma\quad 
t_{\Phi_{1 2}}^{(+)}=1\comma\quad 
t_{\Phi_{2 1}}^{(+)}=-1
\comma\quad t_{\Phi_{1 1}}^{(+)}=t_{\Phi_{22}}^{(+)}=0\comma\\
&\quad t_{\Psi_{1\dot{\alpha}}}^{(+)}=t_{\Psi_{\alpha \dot{2}}}^{(+)}=1/2  \comma \quad t_{\Psi_{2\dot{\alpha}}}^{(+)}=t_{\Psi_{\alpha \dot{1}}}^{(+)}=-1/2\period
\end{aligned}
\eeq 
Comparing \eqref{tchilist} with table \ref{tab:TheChargesFundamentalMagnonTable}, one can see that $t_{\mathcal{X}}^{(+)}$ coincides with the eigenvalues of the R-symmetry rotation generator
\begin{equation}
R = \frac{1}{2}(R^{1}{}_{1}-R^{2}{}_{2} - R^{\dot{1}}{}_{\dot{1}}
+R^{\dot{2}}{}_{\dot{2}}) \period
\end{equation}
Therefore, one can express the flavor-dependent weight factor as
\beq
\mathcal{W}_{{\rm{flavor}}} =    
e^{i J_{\mathcal{X}} \varphi} \, e^{i L_{\mathcal{X}} \phi} \, e^{i R_{\mathcal{X}} \theta}=e^{i L_{\mathcal{X}} \phi} \, e^{i R_{\mathcal{X}} (\theta \pm \varphi)}\comma
\eeq
where $\pm$ correspond  to the $+$ and $-$ dressings respectively. In sum, the net effect of the $Z$-marker dressing for the one-particle state is to change  $\theta$ to $\theta + \varphi$ or $\theta -\varphi$ depending on the choice of the dressings.

\subsection{$Z$-marker dressing for two-particle states}
\begin{figure}[t]
\centering
\includegraphics[clip,height=5cm]{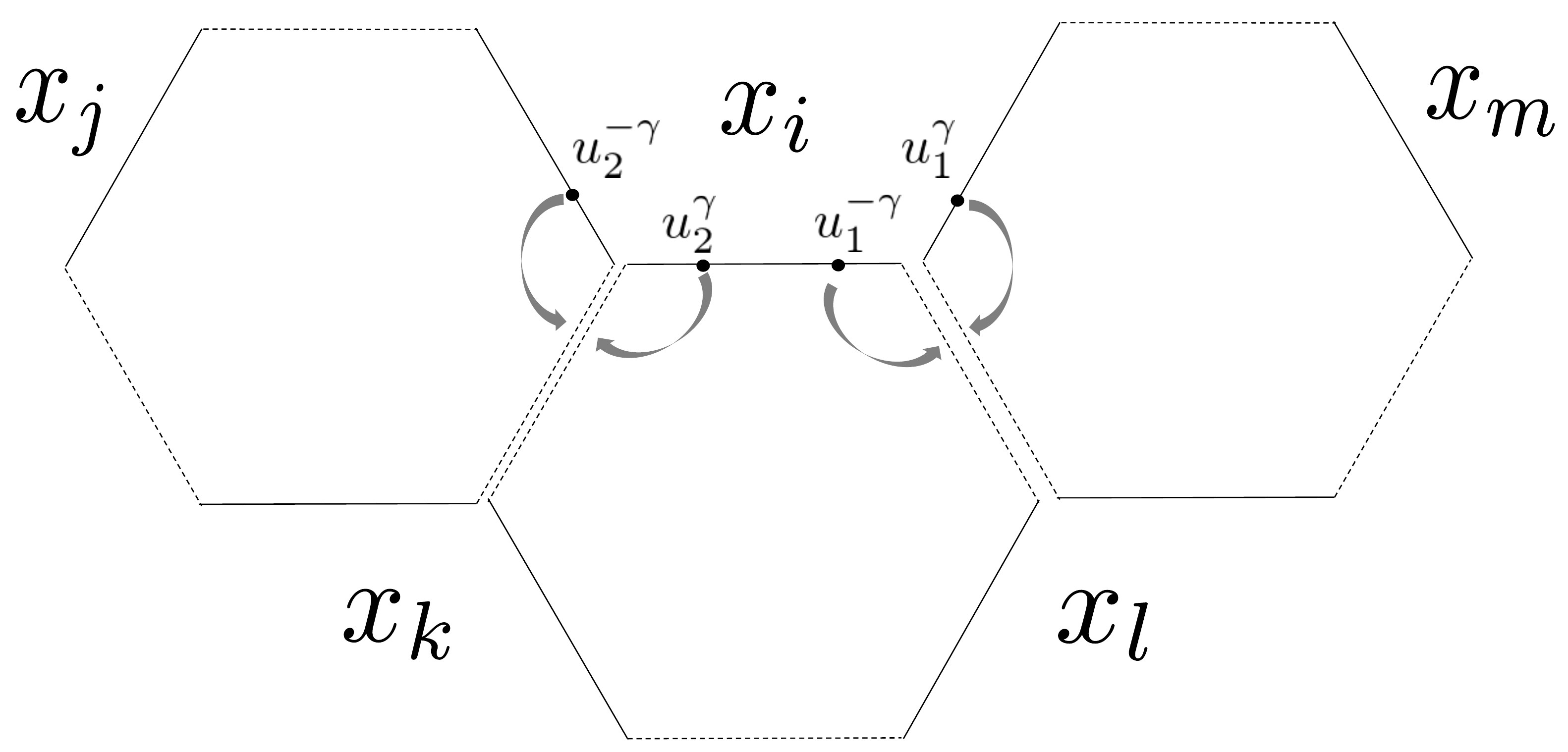}
\caption{The configuration of the mirror particles which produces the two-particle contribution. 
\label{fig:2pt}}
\end{figure}
We now explain how to dress the two-particles 
states (one particle for each edge, see figure \ref{fig:2pt}).
For simplicity, we again consider fundamental magnons only.

The prescription for the two-particle states is a natural generalization of the one for the one-particle states. Namely we propose to dress the state in the following way,
\beq\label{sample}
\underbrace{|Z^{-t_{\mathcal{X}_2}^{(\pm)}}\bar{\mathcal{X}}_2(u_2^{-\gamma})\rangle}_{
\text{left hexagon}}\otimes\underbrace{
|\mathcal{X}_2(u_2^{\gamma})Z^{+t_{\mathcal{X}_2}^{(\pm)}}Z^{-t_{\mathcal{X}_1}^{(\pm)}}
\bar{\mathcal{X}}_1(u_1^{-\gamma})\rangle}_
{\text{middle hexagon}}\otimes
\underbrace{|\mathcal{X}_1(u_1^{\gamma})
Z^{+t_{\mathcal{X}_1}^{(\pm)}}\rangle}_{\text{right hexagon}} \, . 
\eeq
and average over the {\it two} choices, $(\pm)$. (Note that we do not sum over $2\times 2=4$ choices of the signs: The choices of the signs in $t_{\mathcal{X}_1}$ and $t_{\mathcal{X}_2}$ must be correlated.)
For the two-particle state, the effect of the $Z$-markers is twofold: One is to introduce the $J$-charge and change the flavor dependent weight factor. As in the one-particle case, the net effect is to change $\theta_i$ to $\theta_i\pm \varphi_i$. ($\theta_i$ and $\varphi_i$ are the angles for the two channels.) The other is the phase shift induced by moving and removing the $Z$-markers in the state \eqref{sample}. Using the rule given in \cite{BKV}, one obtains
\beq\label{phase}
e^{-i t_{\mathcal{X}_1}^{(\pm)}p(u_2^{\gamma})/2+i t_{\mathcal{X}_2}^{(\pm)} p(u_1^{\gamma})/2}
|\bar{\mathcal{X}}_2(u_2^{-\gamma})
\rangle \otimes
|\mathcal{X}_2(u_2^{\gamma})
\bar{\mathcal{X}}_1(u_1^{-\gamma})\rangle \otimes
|\mathcal{X}_1(u_1^{\gamma})
\rangle \period
\eeq

Using the fact that $t_{\mathcal{X}}^{(+)}$ coincides with the eigenvalue of the $R$ generator, we arrive at the following effective replacement rule, which implements the $Z$-marker dressing: 
\beq
\label{eq:FirstDressing}
\begin{aligned}
\text{$+$ dressing}:&\quad \theta_1 \to 
\theta_1+\varphi_1 -p(u_2^{\gamma})/2\comma\qquad 
\theta_2\to\theta_2+\varphi_2 +p(u_1^{\gamma})/2\comma\\
\text{$-$ dressing}:& \quad \theta_1 \to \theta_1-\varphi_1 +p(u_2^{\gamma})/2\comma\qquad \theta_2\to\theta_2-\varphi_2 -p(u_1^{\gamma})/2\period
\end{aligned}
\eeq
Written more explicitly, the additional phases for scalars in the $+$ dressing are given as follows:
\beq
\begin{aligned}
\mathcal{X}_1(u_1^{ \gamma}):&\quad  
\Phi_{12}(u_1^{ \gamma}) \to 
e^{i(\theta_1+\varphi_1-p(u_2^{\gamma})/2)}\comma
\qquad &&\Phi_{21}(u_1^{\gamma})\to 
e^{-i(\theta_1+\varphi_1-p(u_2^{\gamma})/2)}  \comma\\
\mathcal{X}_2(u_2^{\gamma}):&\quad  
\Phi_{12}(u_2^{\gamma}) \to 
e^{i(\theta_2+\varphi_2+p(u_1^{\gamma})/2)}\comma\qquad 
&&\Phi_{21}(u_2^{\gamma})\to 
e^{-i(\theta_2+\varphi_2+p(u_1^{\gamma})/2)}\period
\end{aligned}
\eeq

As mentioned in the main text, it is convenient for the actual computation to perform the mirror transformation to the middle hexagon and rewrite it as
\beq
|\mathcal{X}_2(u_2^{\gamma})\bar{\mathcal{X}}_1 (u_1^{- \gamma})\rangle\mapsto |\mathcal{X}_1 (u_1^{5 \gamma})\mathcal{X}_2(u_2^{\gamma})\rangle \, .
\eeq
After this rewriting, one can read off the flavor-dependent weight factors just by looking at the charges of what are inside the middle hexagon\fn{Originally the middle hexagon contained $\bar{\mathcal{X}}_1$ and we needed to invert the charge to read off the correct weight factor.}.

The prescription for the $Z$-marker dressing we described here can be straightforwardly generalized to the multi-particle states. It is an interesting future problem to perform the multi-particle computation explicitly and see if the prescription reproduces the correct answer. Also important is to understand the origin of this $Z$-marker prescription. From various circumstantial evidence, we know that it must be related to how the supersymmetry is realized in the hexagon formalism, but it would be nice if we can make it more precise. It would also be desirable to understand it from the viewpoint of the string worldsheet theory\fn{The necessity of the discrete sum is a bit reminiscent of the sum over the spin structure although it cannot be the same thing.}. 

\section{The Bound State-Bound State Mirror S-matrix}\label{ap:S-matrix} 
\label{BoundStateSection}

The matrix part of the hexagon form factor needed to evaluate
the two-particle contribution is given by the 
elements of the mirror bound state $S$-matrix.
The physical bound state $S$-matrix was computed in 
\cite{BoundStateSmatrix} using its Yangian invariance.
In order to compute the necessary mirror bound state $S$-matrix,
we only need to adapt the procedure described in 
\cite{BoundStateSmatrix} to our case. 

\subsection{Basis and Invariant Subsectors} 

In our case, the bound state with index $a$ 
is in the $a$-th anti-symmetric representation of $\mathfrak{su}(2|2)$.
A basis is given by 
\begin{align*}
| \psi_{\alpha_1} \cdots \psi_{\alpha_a} \rangle + \ldots \, , 
\quad \; \;  \; &\quad \quad \quad |\phi_1 \psi_{\alpha_1} \cdots \psi_{\alpha_{a-1}} 
\rangle + \ldots \, ,  \nonumber \\
 |\phi_2 \psi_{\alpha_1} \cdots \psi_{\alpha_{a-1}} 
\rangle+ \ldots \, , 
&\quad \quad \quad |\phi_1 \phi_2 \psi_{\alpha_1} \cdots 
\psi_{\alpha_{a-2}} \rangle + \ldots \, .  \nonumber
\end{align*}

The $S$-matrix is an operator acting in 
the tensor product of two bound states 
with indices $a$ and $b$ and rapidities $u_1$ and $u_2$
respectively. After the action of the $S$-matrix the two 
spaces are interchanged, i.e.
\begin{equation}
S \cdot  \, | {u_1,a} \rangle_i \otimes | {u_2,b} \rangle_j  = (S_{ab})_{i j}^{k l}
(u_1,u_2) \, 
  | {u_2,b} \rangle_k \otimes | {u_1,a} \rangle_l \, , 
\end{equation}
where $i,j,k$ and $l$ denote the basis
elements.

As discuted in \cite{BoundStateSmatrix} for the
physical bound state $S$-matrix and following
the same reasoning, it is possible to show that 
the mirror
bound state $S$-matrix has a
block diagonal structure.  
The $S$-matrix commutes with $\mathfrak{su}(2|2)$ and
the generators of the algebra are the Lorentz rotations $L^{\alpha}_{\; \; \beta},$ 
 the $R$-charge rotations $R^{a}_{\; \; b},$ 
and the supersymmetry generators 
$Q^{\alpha}_{\; \; a}$ and 
$S^{a}_{\;\; \alpha}$. 
In particular, the commutator of $R^{1}_{\; \; 1}$ with the $S$-matrix 
implies that the following  
quantity is conserved  
\begin{equation}
\label{eq:Cconserved}
C_1 = \sharp \, \phi_1^{i}-\sharp \, \phi_2^{i}+\sharp \, \phi_1^{j}
-\sharp \, \phi_2^{j} \, ,  
\end{equation}
where the superscripts $i$ and $j$ refer
to the first and to the second bound state being scattered 
and $\sharp$ means the number of the corresponding fields. 
As a consequence of the form of the basis elements, 
the charge $C_1$ can take the values $C_1 = -2,-1,0,1,2$ and 
one can use it to classify the following invariant subspaces
under the action of the $S$-matrix:  

{\bf{Case I}} 
\begin{itemize}
 
\item $C_1=2$:  
\begin{align*}
|k,l \rangle^{{\rm{Ia}}} = 
| \phi_1 \psi_1^{a - k -1} \psi_2^{k} \rangle \otimes 
| \phi_1 \psi_1^{b - l -1} \psi_2^{l} \rangle \, . 
\end{align*}


\item $C_1=-2$:
\begin{align*}
|k,l \rangle^{{\rm{Ib}}} = 
| \phi_2 \psi_1^{a - k -1} \psi_2^{k} \rangle \otimes 
| \phi_2 \psi_1^{b - l -1} \psi_2^{l} \rangle \, . 
\end{align*}


\end{itemize}

{\bf{Case II}}
\begin{itemize}

\item $C_1 = 1$ : 
\begin{align*}
|k,l \rangle^{{\rm{IIa}}}_1 
&= | \phi_1 \psi_1^{a - k -1} \psi_2^{k} \rangle \otimes 
| \psi_1^{b - l} \psi_2^{l} \rangle \, , \; \;  
|k,l \rangle^{{\rm{IIa}}}_2  = | \psi_1^{a -k} \psi_2^k \rangle
\otimes
|\phi_1 \psi_1^{b -l -1} \psi_2^l \rangle \, , \\
|k,l \rangle^{{\rm{IIa}}}_3  &= | \phi_1 \psi_1^{a-k-1} \psi_2^k 
\rangle \otimes
| \phi_1 \phi_2 \psi_1^{b-l-1} \psi_2^{l-1} \rangle \, , \; \; 
|k,l \rangle^{{\rm{IIa}}}_4  = | \phi_1 \phi_2 \psi_1^{a-k-1} 
\psi_2^{k-1} \rangle  \otimes | \phi_1 \psi_1^{b - l -1} \psi_2^{l} 
\rangle \, . 
\end{align*}

\item $C_1 = -1$ : 
\begin{align*}
|k,l \rangle^{{\rm{IIb}}}_1 
&= | \phi_2 \psi_1^{a - k -1} \psi_2^{k} \rangle \otimes 
| \psi_1^{b - l} \psi_2^{l} \rangle \, , \; \;  
|k,l \rangle^{{\rm{IIb}}}_2  = | \psi_1^{a -k} \psi_2^k \rangle
\otimes
|\phi_2 \psi_1^{b -l -1} \psi_2^l \rangle \, , \\
|k,l \rangle^{{\rm{IIb}}}_3  &= | \phi_2 \psi_1^{a-k-1} \psi_2^k 
\rangle \otimes
| \phi_1 \phi_2 \psi_1^{b-l-1} \psi_2^{l-1} \rangle \, , \; \; 
|k,l \rangle^{{\rm{IIb}}}_4  = | \phi_1 \phi_2 \psi_1^{a-k-1} 
\psi_2^{k-1} \rangle  \otimes | \phi_2 \psi_1^{b - l -1} \psi_2^{l} 
\rangle \, . 
\end{align*}


\end{itemize}

{\bf{Case III}} 
\begin{itemize}

\item $C_1 = 0$
\begin{align*}
|k,l \rangle^{{\rm{III}}}_1 &= 
| \psi_1^{a - k } \psi_2^{k} \rangle \otimes 
| \psi_1^{b - l} \psi_2^{l} \rangle \, ,\\ 
|k,l \rangle^{{\rm{III}}}_2 & = 
| \psi_1^{a -k} \psi_2^k \rangle \otimes 
| \phi_1 \phi_2 \psi_1^{b -l -1} 
\psi_2^{l-1} \rangle \, , \; \; 
|k,l \rangle^{{\rm{III}}}_3  = | \phi_1 \phi_2 \psi_1^{a-k-1}
\psi_2^{k-1} \rangle \otimes 
|\psi_1^{b-l} \psi_2^{l} \rangle \, , \\
|k,l \rangle^{{\rm{III}}}_4 & = 
| \phi_1 \phi_2 \psi_1^{a-k-1} \psi_2^{k-1} \rangle  
\otimes | \phi_1 \phi_2 \psi_1^{b - l -1} \psi_2^{l-1} \rangle 
\, , \\
|k,l \rangle^{{\rm{III}}}_5 & = | 
\phi_1 \psi_1^{a-k-1} \psi_2^k \rangle \otimes 
| \phi_2  \psi_1^{b-l} \psi_2^{l-1} \rangle \, , \; \;  
|k,l \rangle^{{\rm{III}}}_6 = 
| \phi_2  \psi_1^{a-k} \psi_2^{k-1} \rangle \otimes   
| \phi_1 \psi_1^{b - l -1} \psi_2^{l} \rangle \, .  
\end{align*}


\end{itemize}

In addition to the classification of the invariant 
subspaces above, it is possible to extract further 
constraints from the 
conservation of $C_1$ given in 
(\ref{eq:Cconserved}), the commutation of the $S$-matrix with $L^{1}_{\;\; 1}$ and the 
conservation of the bound state indices. 
After a few manipulations, one can show that the 
following quantities are conserved:
\begin{equation}
\begin{aligned}
C_2 & = 2 \, \sharp \phi_2^i + 2 \, \sharp \phi_2^j
+ \sharp \, \psi_1^{i} + \sharp \, \psi_2^{i}+\sharp \, \psi_1^{j}
+ \sharp \, \psi_2^{j} \, , \\
C_3 & = \sharp \, \phi_2^{i} + \sharp \, \phi_2^{j}+\sharp \, 
\psi_2^{i}
+ \sharp \, \psi_2^{j} \, .   \\
\end{aligned}
\end{equation}
Considering $\phi_2$ as a composite state of two fermions 
\cite{BeisertSMatrix1}, the two conserved quantities above
imply the conservation of the total number of 
fermions and of $\psi_2$. 
Thus the $S$-matrix can be written in the form 
($N = k +l$):

{\bf{Case Ia and Ib}} 
\begin{equation}
S \cdot |k,l \rangle^{{\rm{I}}} = 
\sum_{n=0}^{N} \, H^{k,l}_{n} \, |N-n,n \rangle^{{\rm{I}}} \, ,  
\end{equation}

{\bf{Case IIa and IIb}}
\begin{equation}
S \cdot |k,l \rangle_i^{{\rm{II}}} = 
\sum_{n=0}^{N} \, Y^{k,l,j}_{n,i} \, 
|N-n,n \rangle_j^{{\rm{II}}} \, ,   
\end{equation}

{\bf{Case III}}
\begin{equation}
S \cdot |k,l \rangle_i^{{\rm{III}}} = \sum_{n=0}^{N} 
Z^{k,l,j}_{n,i} |N-n,n \rangle_j^{{\rm{III}}} \, .   
\end{equation}

\subsection{Hybrid conventions 
and the action of the Yangian}

The hexagon form factor was derived in \cite{BKV}
and a ``hybrid'' convention was used for 
the excitations in order for the 
$S$-matrix to match the string frame one.
In the ``hybrid'' convention the 
action of the fermionic generators of $\mathfrak{su}(2|2)$ on the
fundamental particles have non-standard 
$Z$-markers and 
are of the form
\begin{equation}
\begin{aligned}
Q^{\alpha}_{\; \; a}
|\phi^{b} \rangle &= a \delta^{b}_{a} \, 
| Z^{\frac{1}{2}}  \psi^{\alpha} \rangle \, , 
\quad \quad \quad \; \, 
Q^{\alpha}_{\; \; a} |\psi^{\beta} 
\rangle = b \epsilon^{\alpha \beta}
 \epsilon_{a b} \, |Z^{\frac{1}{2}} \phi^{b} \rangle \, ,\\
S^{a}_{\; \; \alpha}
|\phi^{b} \rangle &= 
c \epsilon^{ab} \epsilon_{\alpha \beta} \, 
|Z^{-\frac{1}{2}} \psi^{\beta} \rangle \, , 
\quad \; \; 
S^{a}_{\; \; \alpha} |\psi^{\beta} \rangle = 
d \delta^{\beta}_{\alpha} \, 
|Z^{-\frac{1}{2}} \phi^{a} \rangle \, ,\\
\end{aligned}
\end{equation}
with
\begin{equation}
a = \sqrt{g} \gamma \, , \quad 
b = \frac{\sqrt{g}}{\gamma}\left(1-\frac{x^{+}}{x^{-}}
\right) \, , \quad c =  \frac{i \sqrt{g} \gamma}{x^{+}}
\, , \quad d = \frac{\sqrt{g} x^{+}}{i \gamma}
\left(1 - \frac{x^{-}}{x^{+}} \right) \, , 
\end{equation}
and 
\begin{equation}
\gamma = \left( \frac{x^{+}}{x^{-}} \right)^{\frac{1}{4}}
\sqrt{i(x^{-}-x^{+})} \, .
\end{equation}
Note that these transformations are different 
both from the 
spin chain frame transformations of 
\cite{BeisertSMatrix1,BeisertSMatrix2} 
and from the 
string frame transformations of 
\cite{StringFrame,StringFrame2}.
In our conventions for the 
calculations, when a fermionic 
generator acts on a bound state 
basis element we move all 
the $Z$-markers to the right 
of all excitations by using
$Z \chi = e^{- i p} \chi Z$ and then
we delete them. 

The symmetry algebra of the $\mathfrak{su}(2|2)$ fundamental
$S$-matrix was determined in \cite{BeisertYangian}
as the Yangian of the centrally extended $\mathfrak{su}(2|2)$ 
superalgebra, see for example \cite{LecturesYangian}
for an introduction to the Yangian symmetry. This symmetry algebra was used in \cite{BoundStateSmatrix}
to find a closed expression for the physical bound state 
$S$-matrix and we will adapt their construction 
here to compute the mirror bound state $S$-matrix. 
The invariance of the $S$-matrix under the Yangian 
is expressed as 
\begin{equation}
[ \,  \Delta( \mathfrak{J}^A) , \, S \, ] = 0 \; ,
\end{equation}
where $\Delta$ is the coproduct 
of the Yangian algebra and $
\mathfrak{J}^A$ are 
the Yangian generators.

In what follows, we will need 
the coproducts of some level 0 and  1
Yangian generators. The level 1 generators 
are going to be denoted by a hat. 
For completeness, we are going to give 
explicitly formulas for 
a few coproducts in the  
``hybrid'' convention that we are 
going to use.
We have for example for level 0 generators: 
\begin{equation}
\Delta(L^{2}_{\; \; 1}) = L^{2}_{\; \; 1} 
\otimes I + I \otimes L^{2}_{\; \; 1} \, , 
\quad \Delta(Q^{2}_{\; \; 2}) = 
Q^{2}_{\; \; 2} \otimes U^{\frac{1}{2}} + 
I \otimes Q^{2}_{\; \; 2}
\, . 
\end{equation}
where $U$ is an operator that gives $e^{- i p}$
when acting on a state. We act 
with the coproducts on the 
bound states basis 
from the left and additional signs 
can appear because we are working 
with graded vector spaces. 
Moreover,
for level 1 generators, we have
\begin{eqnarray}
\Delta(\hat{L}^{2}_{\; \; 1}) =i u_1 \,  L^{2}_{\; \; 1}
\otimes I + i u_2 \, I \otimes L^{2}_{\; \; 1} \nonumber 
\hspace{30mm} \\
-\frac{1}{2} \, L^{2}_{\; \; \alpha} \otimes 
L^{\alpha}_{\; \; 1}
+ \frac{1}{2} \, L^{\alpha}_{\; \; 1} \otimes 
L^{2}_{\; \; \alpha} + \frac{1}{2} \, S^{a}_{\; \; 1}
\otimes U^{-\frac{1}{2}} Q^{2}_{\; \; a}
+ \frac{1}{2} \, Q^{2}_{\; \; a} \otimes
 U^{\frac{1}{2}} S^{a}_{\; \; 1} \, , 
\end{eqnarray}
\begin{eqnarray}
\Delta(\hat{Q}^{2}_{\; \; 2}) = i u_1 \, Q^{2}_{\; \; 2}
\otimes U^{\frac{1}{2}} + i u_2 \, I \otimes Q^{2}_{\; \; 2}  
\hspace{30mm} 
\nonumber \\
-\frac{1}{2} \, L^{2}_{\; \; \alpha} 
\otimes Q^{\alpha}_{\; \; 2} 
+ \frac{1}{2} \, Q^{\alpha}_{\; \; 2} 
\otimes U^{\frac{1}{2}}
L^{2}_{\; \; \alpha} -
\frac{1}{2} R^{a}_{\; \; 2}
\otimes Q^{2}_{\; \; a} + \frac{1}{2}
\, Q^{2}_{\; \; a} \otimes U^{\frac{1}{2}} 
R^{a}_{\; \; 2} \nonumber \\
- \frac{1}{4} \, H \otimes Q^{2}_{\; \; 2} +
\frac{1}{4} \, Q^{2}_{\; \; 2} \otimes U^{\frac{1}{2}} 
H 
- \frac{1}{2} C \otimes U S^{1}_{\; \; 1}
+ \frac{1}{2} S^{1}_{\; \; 1} \otimes 
U^{-\frac{1}{2}} C \, , \hspace{5mm} 
\end{eqnarray}
\begin{eqnarray}
\Delta(\hat{S}^{1}_{\; \; 1}) =
i u_1 \, S^{1}_{\; \; 1} \otimes U^{-\frac{1}{2}}
+ i u_2 \, I \otimes S^{1}_{\; \; 1} 
\hspace{30mm} \nonumber \\
+ \frac{1}{2} \, L^{\alpha}_{\; \; 1} 
\otimes S^{1}_{\; \; \alpha} - \frac{1}{2} \, 
S^{1}_{\; \; \alpha} \otimes U^{-\frac{1}{2}} 
L^{\alpha}_{\; \; 1} + \frac{1}{2} \,  
R^{1}_{\; \; a}
\otimes S^{a}_{\; \; 1} - \frac{1}{2} \, 
S^{a}_{\; \; 1} \otimes U^{-\frac{1}{2}} 
R^{1}_{\; \; a} \nonumber \\
+ \frac{1}{4} \, H \otimes S^{1}_{\; \; 1} 
- \frac{1}{4} \, S^{1}_{\; \; 1} \otimes U^{-\frac{1}{2}}
H + \frac{1}{2} \, C^{\dag} \otimes U^{-1} 
Q^{2}_{\; \; 2} - \frac{1}{2} \, Q^{2}_{\; \; 2}
\otimes U^{\frac{1}{2}} C^{\dag} \, . 
\end{eqnarray}
where $u_1$ and $u_2$ are
the rapidities of the first and the second bound state
respectively and $H, C$ and $C^{\dag}$ are the central charges
of the algebra:
\begin{equation}
\begin{aligned}
\{ Q^{\alpha}_{\; \; a}, Q^{\beta}_{\; \; b} \} = 
\epsilon^{\alpha \beta} \epsilon_{a b} C \, ,   
\quad \{ S^{a}_{\; \; \alpha}, S^{b}_{\; \; \beta} \} = 
\epsilon_{\alpha \beta} \epsilon^{a b} C^{\dag} \, , \\
\{ Q^{\alpha}_{\; \; a} , S^{b}_{\; \; \beta} \} = \delta^b_a
L^{\alpha}_{\; \; \beta} + \delta^{\alpha}_{\beta}  
R^{b}_{\; \; a} + \frac{1}{2} \delta_{a}^b \delta^{\alpha}_{\beta} H 
\, . \; \; \; \; \; \; 
\end{aligned}
\end{equation}

\subsection{The computation of 
the mirror bound state $S$-matrix} 

In this subsection, we explicitly 
compute the mirror bound state 
$S$-matrix elements
by adapting and following the computation of 
\cite{BoundStateSmatrix}. 
We consider first the basis elements in Case I, then in Case II and 
finally in Case III. 

\subsubsection{Case I}

The cases Ia and Ib are similar, so we will 
consider only
case Ia here and we will omitted the a for simplicity. 
The first step of the calculation is to express the state
$|k,l \rangle^{I}$ for any $k$ and $l$ as 
products of operators acting on $|0,0 \rangle^{I}$.
Indeed, one can show that
\begin{equation}
\label{eq:Theklfromvacuum}
|k,l \rangle^{I} \propto
[(L^{2}_{\; \;  1} \otimes I)( i \delta u - \Delta( 
L^{1}_{\; \; 1}))]^{k} \, 
[(I \otimes L^{2}_{\; \; 1})
( i \delta u - \Delta( 
L^{1}_{\; \; 1}))]^{l} \, |0,0 \rangle^{I} \, ,
\end{equation}
where we have defined $\delta u = u_1 - u_2$
and one can fix the coefficient of 
proportionality by a direct computation.
The particular combination of operators 
appearing on the right hand size
was chosen because 
one can rewrite it as a product of terms involving 
only coproducts by 
using the following relations
(see (4.5)
of \cite{BoundStateSmatrix} for the analogous
equation for the physical case)
\begin{equation}
\label{eq:relationsCase1}
\begin{aligned}
(L^{2}_{\; \; 1} \otimes I)( i \delta u - \Delta( 
L^{1}_{\; \; 1})) |k,l \rangle^{I} & =
( \Delta(\hat{L}^{2}_{\; \; 1}) -i u_2 
\Delta(L^{2}_{\; \; 1}) - 
\Delta(L^{2}_{\; \; 1})(L^{1}_{\; \; 1} \otimes I))
|k,l \rangle^{I} \, ,\\
(I \otimes L^{2}_{\; \; 1})
(i \delta u - \Delta( 
L^{1}_{\; \; 1})) |k,l \rangle^{I} & =
( - \Delta(\hat{L}^{2}_{\; \; 1}) + i u_1 
\Delta(L^{2}_{\; \; 1}) - 
\Delta(L^{2}_{\; \; 1})(I \otimes L^{1}_{\; \; 1}))
|k,l \rangle^{I} \, , 
\end{aligned} 
\end{equation}
and the equations above can be verified by 
replacing the definitions
of the coproducts given previously. 

Replacing the relations (\ref{eq:relationsCase1}) in
(\ref{eq:Theklfromvacuum}), one gets  
a combination of products of 
coproducts. 
The next step is to apply the $S$-matrix operator
to both sides of (\ref{eq:Theklfromvacuum}).
The left hand side will be precisely 
the wanted Case I mirror bound state $S$-matrix 
and the right hand side can be evaluated 
because the $S$-matrix operator 
commutes with any combination of coproducts
and it will be proportional to the action
of it on the state $|0,0\rangle^{I}$.
Thus to complete the computation, 
one needs to evaluate $S \cdot |0,0\rangle^{I}$.
One way of doing the computation is to again use the 
symmetry properties of the $S$-matrix. 
Consider the state 
\begin{equation}
|0,0 \rangle = |\psi_1^{a} \rangle \otimes 
| \psi_1^{b} \rangle \, , 
\end{equation}
The $S$-matrix acts diagonally in the 
state above as it gives a product of the elements
$D_{12}$, see   
\cite{BeisertSMatrix2, CaetanoFleury}. 
In our conventions this element has the value -1, thus
we have 
\begin{equation}
S \cdot | 0 , 0 \rangle = 
(-1)^{(a b)} \, | 0, 0 \rangle \, .   
\end{equation}
Using in addition the following relations 
\begin{equation}
[ \Delta(S^{1}_{\; \; 1}) \Delta(Q^{2}_{\; \; 2}) ,
S \, ] \, | 0 , 0 \rangle  = 0 \, , 
\end{equation}
and that in our normalization
\begin{equation}
\Delta(S^{1}_{\; \; 1}) \Delta(Q^{2}_{\; \; 2}) 
\, | 0,0 \rangle = g \, i \,  (-1)^a 
e^{- i p(u_2)/2} \left[ \frac{1}{x^{-b}(u_2)} - \frac{1}{x^{+a}(u_1)} 
\right] | 0,0 \rangle^{I} \, , 
\end{equation}
one can show that
\begin{equation}
S \cdot |0,0 \rangle^{I} \equiv D_{ab}(u_1,u_2) 
= (-1)^{(a-1)(b-1)} \, \frac{x^{-a}(u_1)-
x^{+b}(u_2)}{x^{+a}(u_1)-
x^{-b}(u_2)} \, 
e^{i p(u_1)/2} \, e^{- i p(u_2)/2} \, ,  
\end{equation}
where $x^{\pm a} = x(u \pm \frac{i}{2} a)$ with $x$ a Zhukowsky
variable defined by $u/g = x+ 1/x$. 

Collecting all the results above, the final expression  
for the mirror bound state $S$-matrix for Case I
is 
\begin{equation}
S \cdot |k,l \rangle^{{\rm{I}}} = 
\sum_{n=0}^{N} \, H^{k,l}_{n} \, |N-n,n \rangle^{
\rm{I}} \, ,   
\end{equation}
where 
\begin{equation}
\label{eq:TheHfunction}
\begin{aligned}
H^{k,l}_{n} = 
D_{a b}(u_1,u_2) 
\times
\frac{ \prod_{m_1=1}^n m_1 \,  
\prod_{m_2=1}^{k+l-n} m_2 }{ 
\prod_{m_3=1}^{k+l} 
[ \,  i  \delta u +(\frac{a+b}{2}-m_3)\,] \,
\prod_{m_4=1}^k m_4 \, \prod_{m_5=1}^l m_5} \times 
 \hspace{22mm} \\ 
\sum_{m=0}^k \hspace{-0.4mm}
\binom{k}{k-m} \hspace{-0.4mm} \binom{l}{n-m}
\hspace{-0.4mm}
\prod_{p=1}^{m} c^+(p) \hspace{-1mm} 
\prod_{p=1-m}^{l-n}  \hspace{-1.2mm}  
c^-(p) \hspace{-0.4mm} 
\prod_{p=1}^{k-m} \hspace{-0.4mm}
d\left(\frac{k-p+2}{2}\right) \hspace{-0.4mm} 
\prod_{p=1}^{n-m}  \hspace{-0.5mm} 
\tilde{d}\left(\frac{k+l-m-p+2}{2}\right) ,
\end{aligned}
\end{equation}
and we have used the definitions
\begin{equation}
\begin{aligned}
c^+(t) &= 
i \delta u
-\frac{(a-b)}{2}+t-1 \, , \quad  
\quad d(t)= -(a+1-2t) \, , \\
c^-(t)&= i \delta u+ \frac{(a-b)}{2}+t-1 \, , 
\quad \quad  
\tilde{d}(t)= -(b+1-2t) \, .
\end{aligned}
\end{equation}
Note that in our normalization conventions 
\begin{equation}
\begin{aligned}
L^{2}_{\; \; 1}  \cdot | \phi_1 \psi_1^{a - k -1} \psi_2^{k} \rangle 
&= (k+1) | \phi_1 \psi_1^{a - k -2} \psi_2^{k+1} \rangle \, , \\
 L^{1}_{\; \; 2}  \cdot | \phi_1 \psi_1^{a - k -1} \psi_2^{k} \rangle 
&= (a-k) | \phi_1 \psi_1^{a - k} \psi_2^{k-1} \rangle \, .
\end{aligned}
\end{equation}
The formula (\ref{eq:TheHfunction}) is an adaptation
of the formula (4.11) of \cite{BoundStateSmatrix}
and we have tested it for many values of
the bound state indices $a$ and $b$.

\subsubsection{Case II}

Analogously to the previous case, 
we are not going 
to distinguish Case IIa and IIb as the computation for both
cases are similar. We will consider Case IIa and omit the a 
for simplicity.     
The Case II mirror bound state $S$-matrix 
can also be fixed by using its Yangian invariance
and by using the result of Case I. 
Recall 
that by symmetries arguments   
the Case II $S$-matrix can be written in the form 
\begin{equation}
S \cdot |k,l \rangle_i^{\rm{II}} = \sum_{n=0}^{N}
Y^{k,l,j}_{n,i} |N-n,n \rangle_j^{
\rm{II}} \, .  
\end{equation}

The idea of 
\cite{BoundStateSmatrix} 
is to derive a set of equations 
involving both the Case II and Case I 
$S$-matrices. In matrix form, the equations 
takes the form 
\begin{equation}
A \cdot  \mathbf{Y}^{k,l}_n
= B^{+} H^{k+1,l-1}_{n} + B^{-} H^{k-1,l+1}_{n}
+ B H^{k,l}_n \, ,  
\end{equation}
where
\begin{equation}
    \mathbf{Y}^{k,l}_n = \left(
      \begin{array}{cccc}
        Y^{k,l,1}_{n,1} & Y^{k,l,1}_{n,2}  & Y^{k,l,1}_{n,3} & Y^{k,l,1}_{n,4}   \\
       Y^{k,l,2}_{n,1} & Y^{k,l,2}_{n,2}  & Y^{k,l,2}_{n,3} & Y^{k,l,2}_{n,4}   \\
Y^{k,l,3}_{n,1} & Y^{k,l,3}_{n,2}  & Y^{k,l,3}_{n,3} & Y^{k,l,3}_{n,4}   \\
Y^{k,l,4}_{n,1} & Y^{k,l,4}_{n,2}  & Y^{k,l,4}_{n,3} & Y^{k,l,4}_{n,4}   
      \end{array} \right) \, , 
      \end{equation}
and $A, B^{+},B^{-}$ and $B$ are matrices.
In what follows we are going to derive the entries
of these matrices. In this work, we are interested
in the one-loop result, so we are not 
going to give a closed expression for the inverse
matrix $A^{-1}$ and for $Y^{k,l,j}_{n,i}$
valid at any value of the coupling constant. 
Instead, one can solve the matrix equation at the necessary 
order in $g^2$.

The first line of the matrix equation ($i=1,2,3,4$) 
is obtained from 
\begin{equation}
^{\rm{I}}\langle N-n,n| \, \Delta(Q^{2}_{\; \; 2}) 
\, S \, |k,l \rangle^{\rm{II}}_i
= \; ^{\rm{I}}\langle N-n,n| \, S \, \Delta(Q^{2}_{\; \; 2}) 
 \, |k,l \rangle^{\rm{II}}_i \, , 
\end{equation}
Similarly, the second line is obtained from    
\begin{equation}
^{\rm{I}}\langle N-n,n| \, \Delta(S^{1}_{\; \; 1}) 
\, S \, |k,l \rangle^{\rm{II}}_i
= \; ^{\rm{I}}\langle N-n,n| \, S \, 
\Delta(S^{1}_{\; \; 1}) 
 \, |k,l \rangle^{\rm{II}}_i \, . 
\end{equation}
Note that in the equations above only
the coproduct of level 0 generators appeared.
The remaining linearly independent  
equations are derived 
using also the 
coproducts of level 1 generators.  
Consider the following combination
of coproducts in our normalization
\begin{eqnarray}
\label{eq:TheGeneratorsTilda}
\tilde{S}^{1}_{\; \; 1} 
= \Delta(\hat{S}^{1}_{\; \; 1}) + a_1 \, \Delta(
\hat{L}^{2}_{\; \; 1}) \,  
\Delta(S^{1}_{\; \; 2}) + a_2 \, 
\Delta(
L^{2}_{\; \; 1}) \,  
\Delta(S^{1}_{\; \; 2}) \, , \quad {\rm{and}} 
\hspace{39mm} 
\end{eqnarray}
\begin{eqnarray}
\tilde{Q}^{2}_{\; \; 2} =
\Delta(\hat{Q}^{2}_{\; \; 2})
- a_1 \, \Delta(
\hat{L}^{2}_{\; \; 1}) \,  
\Delta(Q^{1}_{\; \; 2}) - a_2 \, 
\Delta(
L^{2}_{\; \; 1}) \,  
\Delta(Q^{1}_{\; \; 2}) \, , \quad {\rm{with}}
\hspace{35mm} \nonumber
\\
a_1 = \frac{2}{a+b-2(N+1-i \delta u)}, 
\quad a_2
= \frac{a-b+ 2(N -2 n - i u_1 - i u_2)}{
2(a+b) - 4(N+1- i \delta u)} \, .\nonumber 
\end{eqnarray}

The third and forth line of the matrix 
equations are obtained from 
\begin{equation}
^{\rm{I}}\langle N-n,n| \, \tilde{Q}^{2}_{\; \; 2} \, 
\, S \, |k,l \rangle^{\rm{II}}_i
= \; ^{\rm{I}}\langle N-n,n| \, S \, 
\tilde{Q}^{2}_{\; \; 2} \, 
 \, |k,l \rangle^{\rm{II}}_i \, , 
\end{equation}
and
\begin{equation}
^{\rm{I}}\langle N-n,n| \, \tilde{S}^{1}_{\; \; 1} 
\, S \, |k,l \rangle^{\rm{II}}_i
= \; ^{\rm{I}}\langle N-n,n| \, S \, 
\tilde{S}^{1}_{\; \; 1} 
 \, |k,l \rangle^{\rm{II}}_i \, . 
\end{equation}

The procedure to obtain the values of the coefficients  
$a_1$ and $a_2$ is to impose that the left hand side
of the above equations only contain $Y^{k,l,j}_{n,i}$.
Notice that for general $a_1$ and $a_2$ terms of the 
form $Y^{k,l,j}_{n+1,i}$ can also appear for example,
see \cite{BoundStateSmatrix} for more details.

\subsubsection{Case III}

Recall that the Case III 
$S$-matrix can be written in the form  
\begin{equation}
S \cdot |k,l \rangle_i^{{\rm{III}}} = \sum_{n=0}^{N} 
Z^{k,l,j}_{n,i} |N-n,n \rangle_j^{{\rm{III}}} \, .   
\end{equation}

The way of computing it is to use again the symmetry algebra 
of the $S$-matrix and relate the Case III to the already known results for
Case II and Case I, see \cite{BoundStateSmatrix}. 
In this case, only the coproducts involving level 0 generators are necessary.
The key observation is that acting in any 
Case III basis element with
the coproducts of $Q^{2}_{\; \; 2}, Q^{1}_{\; \; 1}, S^{1}_{\; \; 1}$ and 
$S^{2}_{\; \; 2}$ one gets a result proportional to a basis element 
of Case II. In other words, using some of the relations
\begin{equation}
\begin{aligned}
^{\rm{II}}_i \langle N -n, n | [ \Delta(Q^{1}_{\; \; 1}), S ] | k , l \rangle^{{\rm{III}}}_j
=0 \, , \quad  ^{\rm{II}}_i \langle N -n, n | [ \Delta(Q^{2}_{\; \; 2}), S ] | k , l \rangle^{{\rm{III}}}_j =0 \, , \\
^{\rm{II}}_i \langle N -n, n | [ \Delta(S^{1}_{\; \; 1}), S ] | k , l \rangle^{{\rm{III}}}_j
=0 \, , \quad  ^{\rm{II}}_i \langle N -n, n | [ \Delta(S^{2}_{\; \; 2}), S ] | k , l \rangle^{{\rm{III}}}_j =0 \, ,
\end{aligned}
\end{equation} 
one can select a set of linearly independent equations 
to write a matrix equation for the elements of the Case III $S$-matrix 
as function of the elements of the Case II. 
The solution can be expanded up to the necessary order in powers of $g^2$.

\section{Weak Coupling Expansions}\label{WeakCouplingExpansions} 

In this appendix, we defined and
perform the weak coupling expansion of
necessary quantities for the computation
of the two-particle contribution 
at order $g^2$. The fused dynamical part
of the hexagon form factor is given by
\begin{equation}
h_{ab}(u,v) = \prod_{k= - \frac{a-1}{2}}^{\frac{a-1}{2}} 
\, 
\prod_{l= - \frac{b-1}{2}}^{\frac{b-1}{2}} 
\, h(u^{[2k]},v^{[2l]}) \, . 
\end{equation}
In fact, we will need the mirror rotated 
fused dynamical part. To evaluate this quantity,
it is necessary to compute the mirror 
rotated dressing phase \cite{CrossingI,CrossingII}. 
Different results 
are obtained depending
on the order, i.e. the processes of fusion 
and crossing do not commute for the dressing phase. 
Here,
the correct procedure is to first fuse and then
crossing and we get at order $g^0$:

\begin{equation}
\sigma_{a b}(u^{\gamma}, v^{\gamma}) = 
\frac{\Gamma[1- \frac{a}{2} + i u] 
\Gamma[1 + \frac{a-b}{2}
-i(u-v)] \Gamma[1+ \frac{b}{2}-iv]}{
\Gamma[1+ \frac{a}{2} - i u]
\Gamma[1 + \frac{b-a}{2}
+i(u-v)] \Gamma[1-\frac{b}{2}+i v]} \, ,
\end{equation}

\begin{equation}
\sigma_{a b}(u^{\gamma}, v^{-\gamma}) = 
\frac{\Gamma[1+ \frac{a}{2} - i u] 
\Gamma[1 - \frac{a+b}{2}
+i(u-v)] \Gamma[1+ \frac{b}{2}+iv]}{
\Gamma[1- \frac{a}{2} + i u]
\Gamma[1 + \frac{a+b}{2}
-i(u-v)] \Gamma[1-\frac{b}{2}-i v]} \, .
\end{equation}

We have at order $g^2$ 
\begin{eqnarray}
h_{ab}(u^{\gamma},v^{\gamma}) =
\frac{g^2}{\sigma_{a b}(u^{\gamma}, v^{\gamma})}
\frac{(\frac{(a+b)^2}{4}+(u-v)^2)}{(\frac{a^2}{4}+u^2)
(\frac{b^2}{4}+v^2)} \times \hspace{50mm} \nonumber\\
\frac{\Gamma[-\frac{a}{2} - i u]
\Gamma[\frac{a+b}{2}-i(u-v)]
\Gamma[-\frac{a+b}{2}+i(u-v)]\Gamma[\frac{b}{2}-i v]}{
\Gamma[\frac{a}{2} - i u]\Gamma[\frac{b-a}{2}-i(u-v)]
\Gamma[\frac{b-a}{2}+i(u-v)]\Gamma[-\frac{b}{2}-i v]} \, .
\end{eqnarray}

The momentum and the exponential 
of the energy for mirror 
bound states are

\begin{equation}
\label{eq:weakmomentum}
\begin{aligned}
\tilde{p}_a(u) &= u + O(g^2) \, ,  \\
e^{- \tilde{E}_a(u)} &= \frac{g^2}{(u^2+ \frac{a^2}{4})}
+ O(g^4) \, .
\end{aligned}
\end{equation}

Finally the measure for mirror bound states is 
\begin{equation}
\mu_{a}(u^{\gamma}) = \frac{a g^2}{
(u^2 + \frac{a^2}{4})^{2}} +
O(g^4) \, . 
\end{equation}

\section{The Integrand and the Integral 
for the Two-Particle Contribution}\label{TheIntegrand}
 \subsection{Explicit form of the integrand}
In this section, we are going to write down the complete two-particle integrand.
We will use the mirror bound state $S$-matrix elements 
of Appendix \ref{BoundStateSection} and the $Z$-markers
prescription of Appendix \ref{ZmarkersSection}.  Consider the  figure 
\ref{fig:2pt}. We are going to compute the hexagon form factor of the 
middle hexagon by applying mirror tranformations to the particle
1 to get $u_1^{5 \gamma}$ (Of couse, this is not necessary 
and the result of the integral must be the same 
if one works with $u_1^{-\gamma}$). 
The hexagon form factors of the 
left and right hexagons of the 
figure are trivial because they only have
one bound state. Their values is a product
of one particle hexagon form factors and 
they can give only a non-trivial sign.
Using the important identity
\begin{equation}
h(u^{4 \gamma},v)=\frac{1}{h(v,u)} \, ,
\end{equation}
where $h(u,v)$ is the dynamical part 
of the hexagon form factor, the two-particle
contribution is 
\begin{eqnarray}
\label{eq:TheIntegral}
\mathcal{M}^{(2)} (z_1,z_2,\alpha_1,\alpha_2) = 
\int \frac{d u_1}{2 \pi} 
\frac{d u_2}{2 \pi} \sum_{a,b}
\frac{\mu_a(u_1^{\gamma})  
\mu_b(u_2^{\gamma}) }{h_{b a}(u_2^{\gamma},u_1^{\gamma})} 
\, e^{- 2 i \tilde{p}_a(u_1) {\rm{log}}|z_1|}
e^{- 2 i \tilde{p}_b(u_2) {\rm{log}}|z_2|}\, 
\mathcal{F}_{ab} \, . 
\end{eqnarray}
In the expression above, $\mu_a(u)$ are the measures,
the exponentials are the flavor independent 
part of the weight factors,
$z_1$ and $z_2$ are the relevant
cross-ratios for the right edge and 
the left edge respectively.
$\mathcal{F}_{ab}$ is essentially
the matrix
part of the middle hexagon form factor which
contains the interaction between the two mirrors
particles and its expression will
be given below. 
The flavor dependent part of the weight factors
will be written in terms of the angles: 
\beq
e^{i\phi_i}=\sqrt{\frac{z_i}{\bar{z}_i}}\comma\quad 
e^{i\theta_i}=\sqrt{\frac{\alpha_i}{\bar{\alpha}_i}}\comma 
\qquad e^{i\varphi_i}=
\sqrt{\frac{\alpha_i\bar{\alpha}_i}{z_i\bar{z}_i}}
\comma
\eeq
with $\alpha_i$ the $R$-charge cross-ratios. 

The matrix part is a sum of several terms
coming from the different elements of 
the mirror bound state $S$-matrix. Before 
considering all the cases, let us first make a list
of all contributing signs: 


\begin{itemize}

\item There is a factor of  
$(-1)^{F_1} (-1)^{F_2}$ where
$F_i$ is the fermion number of each state.
These signs appear because in the string
frame the one-particle hexagon form
factor differs by a factor of $-i$
for bosonic and fermionic indices 
\cite{CaetanoFleury}.

\item The factor $(-1)^{\mathfrak{f}}$ of the
middle hexagon gives $(-1)^{F_1 F_2}$.

\item The mirror transformations of 
the particle 1 give $(-1)^{a}$

\item The left and right hexagons form factors and
the one particle hexagon form factors 
contractions gives $(-1)^{b}$. 

\end{itemize}

Using the $Z$-markers prescription for the two-particle
case, the $S$-matrix of appendix 
\ref{BoundStateSection} and recalling 
that one has to average the result of
both
the $+$ dressing and the $-$ dressing, 
the matrix part 
$\mathcal{F}_{ab}(u_1,u_2)$ is the sum of the following
terms 

\begin{itemize}

\item {\bf{Case Ia and Ib}}
\begin{eqnarray} 
2 \,  (-1)^{(a-1)(b-1)} \,  
{\rm{cos}}(\theta_1+\theta_2) \, 
{\rm{cos}}(\varphi_1+\varphi_2 +\frac{p(u_1^{\gamma})-
p(u_2^{\gamma})}{2}) \times \nonumber 
\hspace{30mm} \\
\sum_{k=0}^{a -1} \sum_{l=0}^{b-1} 
e^{i \phi_1 (a - 2k -1)}
e^{i \phi_2 (b - 2l -1)} H^{k,l}_{k} \, .
\nonumber
\end{eqnarray}

\item {\bf{Case IIa and Case IIb}}

\begin{eqnarray}
- 2 (-1)^{(a-1)b} {\rm{cos}}(\theta_1) {\rm{cos}}
(\varphi_1 - \frac{p(u_2^{\gamma})}{2})
\sum_{k=0}^{a -1} 
\sum_{l=0}^{b} e^{i \phi_1(a-2k-1)}e^{i \phi_2 
(b-2l)} Y^{k,l,2}_{k,1} \nonumber \\
- 2 (-1)^{a(b-1)} {\rm{cos}}(\theta_2) 
{\rm{cos}}(\varphi_2+ \frac{p(u_1^{\gamma})}{2})
\sum_{k=0}^{a} 
\sum_{l=0}^{b-1} e^{i \phi_1 (a-2k)}
e^{i \phi_2 (b-2l-1)} 
Y^{k,l,1}_{k,2} \nonumber \\
- 2 (-1)^{(a-1)b} {\rm{cos}}(\theta_1) {\rm{cos}}
(\varphi_1 - \frac{p(u_2^{\gamma})}{2})
\sum_{k=0}^{a -1} 
\sum_{l=1}^{b-1} e^{i \phi_1(a-2k-1)}e^{i \phi_2 
(b-2l)} Y^{k,l,4}_{k,3} \nonumber \\
- 2 (-1)^{a(b-1)} {\rm{cos}}(\theta_2) 
{\rm{cos}}(\varphi_2+ \frac{p(u_1^{\gamma})}{2})
\sum_{k=1}^{a-1} 
\sum_{l=0}^{b-1} e^{i \phi_1 (a-2k)}
e^{i \phi_2 (b-2l-1)} 
Y^{k,l,3}_{k,4} \nonumber 
\end{eqnarray}

\item  {\bf{Case III}}
\begin{eqnarray}
(-1)^{a b} \sum_{k=0}^{a} 
\sum_{l=0}^{b} e^{i \phi_1 (a - 2k)}
e^{i \phi_2 (b - 2l)} Z^{k,l,1}_{k,1} \, + \, 
(-1)^{a b} \sum_{k=0}^{a} 
\sum_{l=1}^{b-1} e^{i \phi_1 (a - 2k)}
e^{i \phi_2 (b - 2l)} Z^{k,l,3}_{k,2}  \nonumber 
\\
+ \, 
(-1)^{a b} \sum_{k=1}^{a-1} 
\sum_{l=0}^{b} e^{i \phi_1 (a - 2k)}
e^{i \phi_2 (b - 2l)} Z^{k,l,2}_{k,3} \, +
\, (-1)^{a b} \sum_{k=1}^{a-1} 
\sum_{l=1}^{b-1} e^{i \phi_1 (a - 2k)}
e^{i \phi_2 (b - 2l)} Z^{k,l,4}_{k,4} \nonumber 
\\
+ (-1)^{(a-1)(b-1)} e^{i \theta_1}
e^{-i \theta_2} 
{\rm{cos}}(\varphi_1-\varphi_2-\frac{
p(u_2^{\gamma})+p(u_1^{\gamma})}{2})
\sum_{k=0}^{a-1} 
\sum_{l=1}^{b} e^{i \phi_1 (a - 2k-1)}
e^{i \phi_2 (b - 2l+1)} Z^{k,l,6}_{k,5} \nonumber \\
+ (-1)^{(a-1)(b-1)} e^{-i \theta_1}
e^{i \theta_2} 
{\rm{cos}}(\varphi_1-\varphi_2-\frac{
p(u_2^{\gamma})+p(u_1^{\gamma})}{2})
\sum_{k=1}^{a} 
\sum_{l=0}^{b-1} 
e^{i \phi_1 (a - 2k+1)}
e^{i \phi_2 (b - 2l-1)} Z^{k,l,5}_{k,6} \nonumber
\end{eqnarray}

\end{itemize}

The complete integrand is the matrix part given
by the sum of all the terms above and the 
other terms in (\ref{eq:TheIntegral}). 
It remains to expand the integrand 
up to desire order. One can use
the weak coupling expansions of the appendix
\ref{WeakCouplingExpansions} to 
get it at order $g^2$.

To acutually perform the integral, we first evaluated them by taking the residues up to certain values of $a$ and $b$. This produces the expansion of the final integral in $z_1$ and $z_2$. 
We then compared that expansion with the expansion of some ansatz that consist of a linear 
combinations of functions 
and fit for the coefficients. The ansatz is given by a linear combination of one-loop conformal integrals with the arguments being various
possible cross ratios that one can have for the five-point function\footnote{We are 
obligated to 
P. Vieira for helping us to compute
the integral and suggesting the basis of functions.}. 
We also checked that the result is correct by 
computing some of the integrals numerically for 
given values of $z_1$ and $z_2$. 
The final result is given in the main text in 
(\ref{eq:twoparticle}). 
\subsection{The $i \epsilon$ prescription}
\begin{figure}[t]
\centering
\begin{minipage}{0.45\hsize}
\includegraphics[clip,height=5cm]{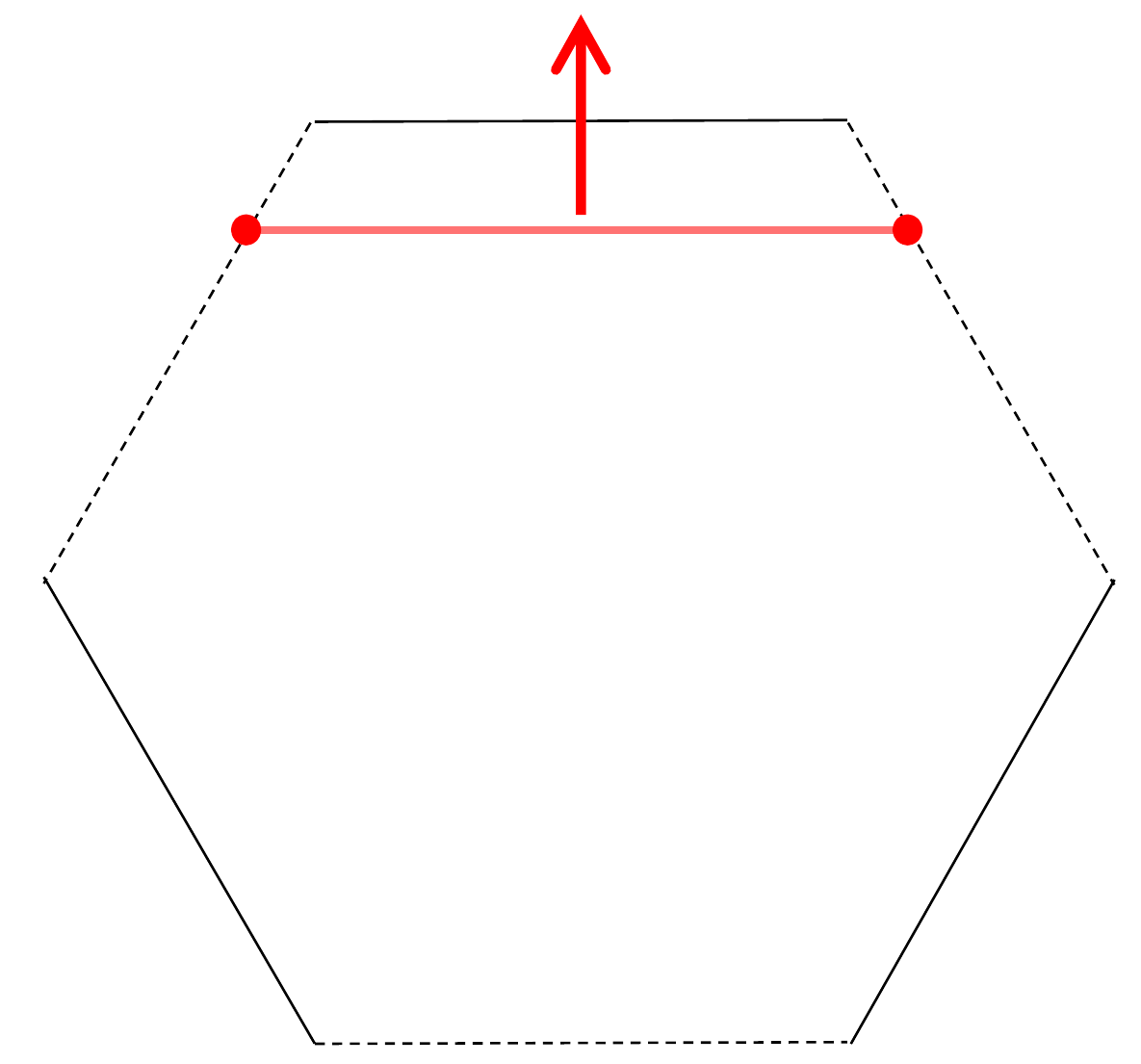}
\end{minipage}
\begin{minipage}{0.45\hsize}
\includegraphics[clip,height=5cm]{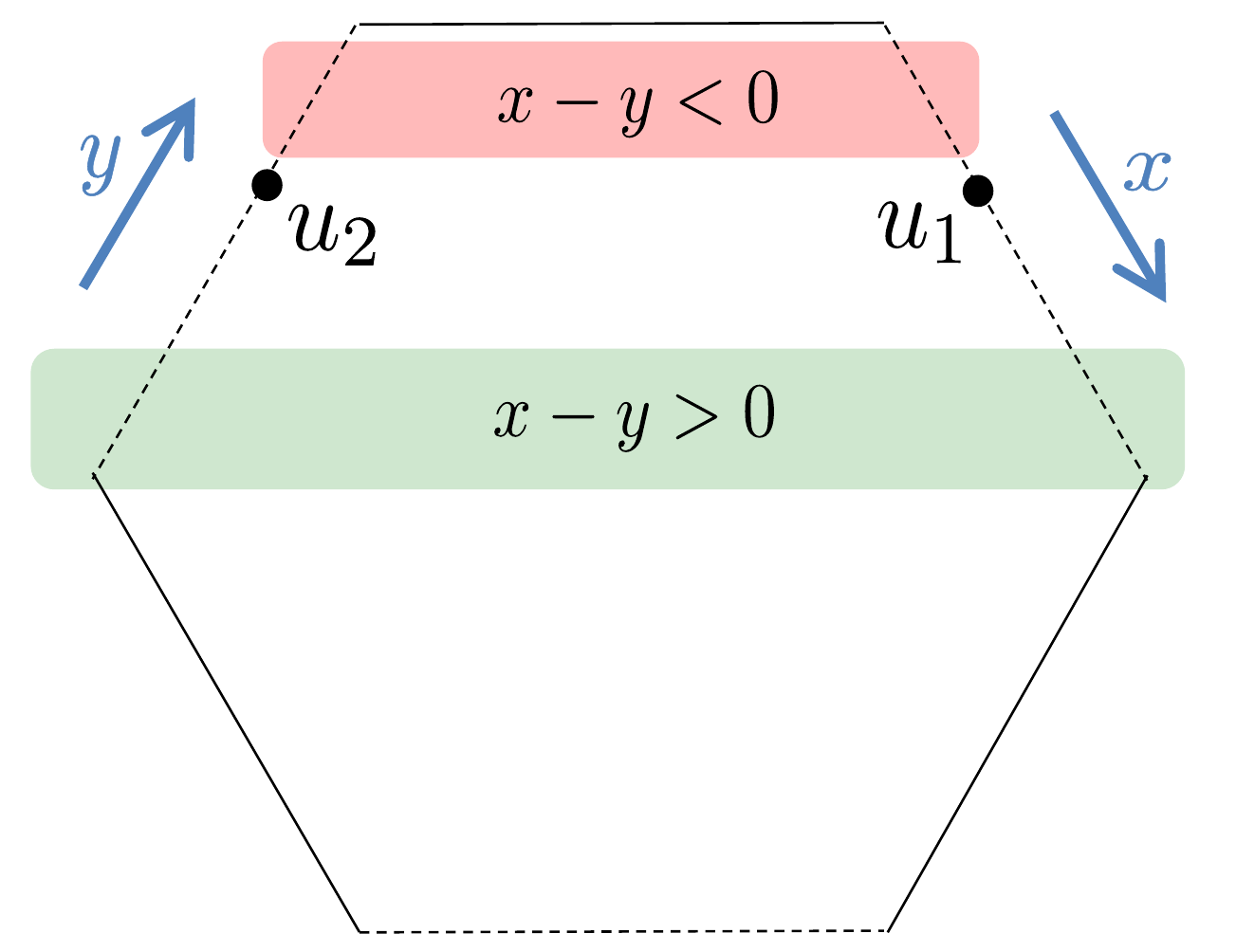}
\end{minipage}
\caption{The kinematical pole and its physical meaning. Left: The kinematical pole arises from the physical process in which the two particles move together in the direction depicted in the figure and decouple from the rest. Right: The coordinates on the mirror edges. $x$ and $y$ are the coordinates on the mirror edges which run from $-\infty$ to $\infty$. The arrows indicate the directions in which these coordinatres increase. The two shaded regioins denote the region $x-y>0$ and $x-y<0$ respectively.}
\label{fig:iepsilon}
\end{figure}

Before closing this appendix, let us make one more comment about the integral. Whenever the bound state indices are the same $(a=b)$, the integral contains a simple pole at $u_1=u_2$, which corresponds to the so-called kinematical singularity. Physically, this singularity represents the IR divergence which arises from the two particles moving together and decoupling from the hexagon (see figure \ref{fig:iepsilon}). Since this pole lies right on top of the integration contour, one has to specify how to avoid it in order to get a meaningful result\fn{For the Pentagon Program for the null polygonal Wilson loops, the correct $i\epsilon$ prescription was discussed in \cite{Pentagonie}. The argument presented here is essentially the same as the one in that paper.}. 

To see what is the correct prescription, it is useful to consider the form factor in the position space rather than in the rapidity (or equivalently in the momentum) space. As usual, the conversion is done by the Fourier transformation,
\beq
\frac{1}{u_1-u_2}\quad \to \quad \int d\tilde{p}_1 d\tilde{p}_2\frac{1}{u_1-u_2 \pm i\epsilon} e^{i\tilde{p}_1 x + i\tilde{p}_2 y}\period
\eeq
Here $x$ and $y$ are the coordinates of the mirror edges which, in our convention, run in the directions depicted in figure \ref{fig:iepsilon}. We also put $\pm i \epsilon$ in the denominator and they correspond to two different ways of avoiding the kinematical pole. Now, using the weak coupling expansion of the momenta $\tilde{p}_i$
given in (\ref{eq:weakmomentum}), one can perform the above integral to get
\beq\label{eq:stepfunction}
\begin{aligned}
\int d\tilde{p}_1 d\tilde{p}_2\frac{1}{u_1-u_2 \pm i\epsilon} e^{i\tilde{p}_1 x + i\tilde{p}_2 y}&\propto \int d \delta \tilde{p} \frac{1}{u_1-u_2 \pm i\epsilon} e^{i\delta\tilde{p} (x-y)}\\
&\propto \Theta (\pm (x-y))\comma
\end{aligned}
\eeq
where $\delta\tilde{p}=(\tilde{p}_1-\tilde{p}_2)/2$ and $\Theta (z)$ is the Heaviside step function. 

The equation \eqref{eq:stepfunction} shows that the effect of the kinematical pole is visible in the region $x-y>0$ if we choose $+i\epsilon$ while it is visible in the region $x-y<0$ if we choose $-i\epsilon$. However, from the figure \ref{fig:iepsilon}, it is clear that the kinematical pole corresponds to the process in which the two particles move together in the region $x-y<0$. Therefore we conclude that the correct choice is $-i\epsilon$. In fact, it is this choice ($-i\epsilon$) that reproduces the perturbative data and, if we choose the other one ($+i \epsilon$), the results would not agree with the perturbation theory.


\begin{thebibliography}{99}
\bibitem{review}
 N.~Beisert {\it et al.},
  ``Review of AdS/CFT Integrability: An Overview,''
  Lett.\ Math.\ Phys.\  {\bf 99}, 3 (2012)
  \arXiv{1012.3982}{1012.3982}.
\bibitem{QuantumSpectralCurve}
  N.~Gromov, V.~Kazakov, S.~Leurent and D.~Volin,
  ``Quantum Spectral Curve for Planar $\mathcal{N} =$ Super-Yang-Mills Theory,''
  Phys.\ Rev.\ Lett.\  {\bf 112} (2014) no.1,  011602
  \arXiv{1305.1939}{1305.1939}.
\bibitem{GromovIntroduction}
  N.~Gromov,
  ``Introduction to the Spectrum of $\mathcal{N}=4$ SYM and the Quantum Spectral Curve,''
  \arXiv{1708.03648}{1708.03648}.

\bibitem{EdenSfrondrini}
  B.~Eden and A.~Sfondrini,
  ``Tessellating cushions: four-point functions in N=4 SYM,''
  \arXiv{1611.05436}{1611.05436}.

\bibitem{FleuryKomatsu}
  T.~Fleury and S.~Komatsu,
  ``Hexagonalization of Correlation Functions,''
  JHEP {\bf 1701} (2017) 130
  \arXiv{1611.05577}{1611.05577}.

\bibitem{BKV} 
  B.~Basso, S.~Komatsu and P.~Vieira,
  ``Structure Constants and Integrable Bootstrap in Planar N=4 SYM Theory,''
  \arXiv{1505.06745}{1505.06745}.


\bibitem{AsymptoticFour}
  B.~Basso, F.~Coronado, S.~Komatsu, H.~T.~Lam, P.~Vieira and D.~l.~Zhong,
  ``Asymptotic Four Point Functions,''
  \arXiv{1701.04462}{1701.04462}.


  
\bibitem{DrukkerPlefka}
  N.~Drukker and J.~Plefka,
  ``The Structure of n-point functions of chiral primary operators in N=4 super Yang-Mills at one-loop,''
  JHEP {\bf 0904} (2009) 001
  \arXiv{0812.3341}{0812.3341}.

\bibitem{nonplanarpaper}
T~Bargheer, J.~Caetano, T.~Fleury, S.~Komatsu and P.~Vieira, ``Handling Handles I: Nonplanar Integrability,'' to appear.

\bibitem{nonplanarpaper2}
T~Bargheer, J.~Caetano, T.~Fleury, S.~Komatsu and P.~Vieira, ``Handling Handles II: Stratification and Data Analysis,'' to appear.

\bibitem{EdenNew}
  B.~Eden, Y.~Jiang, D.~l.~Plat and A.~Sfondrini,
  ``Colour-dressed hexagon tessellations for 
  correlation functions and non-planar corrections,''
  \arXiv{1710.10212}{1710.10212}.


\bibitem{Extremal2} 
  B.~Eden, P.~S.~Howe, C.~Schubert, E.~Sokatchev and P.~C.~West,
  ``Extremal correlators in four-dimensional SCFT,''
  Phys.\ Lett.\ B {\bf 472}, 323 (2000)
  \hep{hep-th/9910150}{hep-th/9910150}.
  
\bibitem{Extremal4} 
  J.~Erdmenger and M.~Perez-Victoria,
  ``Nonrenormalization of next-to-extremal correlators in N=4 SYM and the AdS / CFT correspondence,''
  Phys.\ Rev.\ D {\bf 62}, 045008 (2000)
  \hep{hep-th/9912250}{hep-th/9912250}.

\bibitem{Extremal3}
  B.~U.~Eden, P.~S.~Howe, E.~Sokatchev and P.~C.~West,
  ``Extremal and next-to-extremal n point correlators in four-dimensional SCFT,''
  Phys.\ Lett.\ B {\bf 494} (2000) 141
   \hep{hep-th/0004102}{hep-th/0004102}.

\bibitem{AllThreeLoop}
  D.~Chicherin, J.~Drummond, P.~Heslop and E.~Sokatchev,
  ``All three-loop four-point correlators of half-BPS operators in planar $ \mathcal{N} $ = 4 SYM,''
  JHEP {\bf 1608} (2016) 053
  \arXiv{1512.02926}{1512.02926}.
 



\bibitem{BassoDixon}
  B.~Basso and L.~J.~Dixon,
  ``Gluing Ladder Feynman Diagrams into Fishnets,''
  Phys.\ Rev.\ Lett.\  {\bf 119} (2017) no.7,  071601
  \arXiv{1705.03545}{1705.03545}.
  

 \bibitem{KazakovI}
 O.~Gurdogan and V.~Kazakov,
  ``New Integrable 4D Quantum Field Theories from Strongly Deformed Planar $\mathcal N = $ 4 Supersymmetric Yang-Mills Theory,''
  Phys.\ Rev.\ Lett.\  {\bf 117} (2016) no.20,  201602
   Addendum: [Phys.\ Rev.\ Lett.\  {\bf 117} (2016) no.25,  259903]
  \arXiv{1512.06704}{1512.06704}.


\bibitem{SiegDeformed}
  C.~Sieg and M.~Wilhelm,
  ``On a CFT limit of planar $\gamma_i$-deformed $\mathcal{N}=4$ SYM theory,''
  Phys.\ Lett.\ B {\bf 756} (2016) 118
  \arXiv{1602.05817}{1602.05817}.
 

\bibitem{CaetanoKazakov}
  J.~Caetano, O.~Gurdogan and V.~Kazakov,
  ``Chiral limit of N = 4 SYM and ABJM and integrable Feynman graphs,''
  \arXiv{1612.05895}{1612.05895} [hep-th].
 
\bibitem{FishnetI}
  D.~Chicherin, V.~Kazakov, F.~Loebbert, D.~Müller and D.~l.~Zhong,
  ``Yangian Symmetry for Bi-Scalar Loop Amplitudes,''
  \arXiv{1704.01967}{1704.01967}.

\bibitem{FishnetII}
  N.~Gromov, V.~Kazakov, G.~Korchemsky, S.~Negro and G.~Sizov,
  ``Integrability of Conformal Fishnet Theory,''
  \arXiv{1706.04167}{1706.04167}.
 
\bibitem{FishnetIII}
  D.~Chicherin, V.~Kazakov, F.~Loebbert, D.~Müller and D.~l.~Zhong,
  ``Yangian Symmetry for Fishnet Feynman Graphs,''
  \arXiv{1708.00007}{1708.00007}.
\bibitem{BoundStateSmatrix}
  G.~Arutyunov, M.~de Leeuw and A.~Torrielli,
  ``The Bound State S-Matrix for AdS(5) x S**5 Superstring,''
  Nucl.\ Phys.\ B {\bf 819} (2009) 319
  \arXiv{0902.0183}{0902.0183}.
\bibitem{BeisertSMatrix1}
  N.~Beisert,
  ``The SU(2$|$2) dynamic S-matrix,''
  Adv.\ Theor.\ Math.\ Phys.\  {\bf 12} (2008) 945
  \hep{hep-th/0511082}{hep-th/0511082}.
  \bibitem{BeisertSMatrix2}
  N.~Beisert,
  ``The Analytic Bethe Ansatz for a Chain with Centrally Extended su(2$|$2) Symmetry,''
  J.\ Stat.\ Mech.\  {\bf 0701} (2007) P01017
  \hep{nlin/0610017}{nlin/0610017}.


%
%
%
%
  
  



  

 








 




  

\bibitem{StringFrame}
  G.~Arutyunov, S.~Frolov and M.~Zamaklar,
  ``The Zamolodchikov-Faddeev algebra for 
  AdS(5) x S**5 superstring,''
  JHEP {\bf 0704} (2007) 002
  \hep{hep-th/0612229}{hep-th/0612229}.
  
\bibitem{StringFrame2}
  G.~Arutyunov and S.~Frolov,
  ``Foundations of the $AdS_5 x S^5$ 
  Superstring. Part I,''
  J.\ Phys.\ A {\bf 42} (2009) 254003
  \arXiv{0901.4937}{0901.4937}.
 

 
\bibitem{BeisertYangian}
  N.~Beisert,
  ``The S-matrix of AdS / CFT and Yangian symmetry,''
  PoS SOLVAY {\bf } (2006) 002
  \arXiv{0704.0400}{0704.0400}.
  
\bibitem{LecturesYangian}
  F.~Loebbert,
  ``Lectures on Yangian Symmetry,''
  J.\ Phys.\ A {\bf 49} (2016) no.32,  323002
  \arXiv{1606.02947}{1606.02947}.


\bibitem{CaetanoFleury}
  J.~Caetano and T.~Fleury,
  ``Fermionic Correlators from Integrability,''
  JHEP {\bf 1609} (2016) 010
   \arXiv{1607.02542}{1607.02542}.

  

 
\bibitem{CrossingI}
  G.~Arutyunov and S.~Frolov,
  ``The Dressing Factor and Crossing Equations,''
  J.\ Phys.\ A {\bf 42} (2009) 425401
 \arXiv{0904.4575}{0904.4575}.


\bibitem{CrossingII}
  Z.~Bajnok, A.~Hegedus, R.~A.~Janik and T.~Lukowski,
  ``Five loop Konishi from AdS/CFT,''
  Nucl.\ Phys.\ B {\bf 827} (2010) 426
  \arXiv{0906.4062}{0906.4062}.
 

  



\bibitem{WilsonLoops}
  B.~Basso, A.~Sever and P.~Vieira,
  ``Spacetime and Flux Tube S-Matrices at Finite Coupling for N=4 Supersymmetric Yang-Mills Theory,''
  Phys.\ Rev.\ Lett.\  {\bf 111} (2013) no.9,  091602
 \arXiv{1303.1396}{1303.1396}.
\bibitem{Pentagonie} 
  B.~Basso, A.~Sever and P.~Vieira,
  ``Space-time S-matrix and Flux-tube S-matrix III. The two-particle contributions,''
  JHEP {\bf 1408}, 085 (2014)
  \arXiv{1402.3307}{1402.3307}.

  
\end{thebibliography}
\end{document}